\documentclass[acmtosem]{acmtrans2m}

\usepackage{graphicx}
\usepackage{verbatim}
\usepackage{times}
\usepackage[caption=false,font=footnotesize]{subfig}
\usepackage{algorithmic}
\usepackage{tabularx}

\acmVolume{2}
\acmNumber{3}
\acmYear{01}
\acmMonth{09}

\newcommand{\BibTeX}{{\rm B\kern-.05em{\sc i\kern-.025em b}\kern-.08em
    T\kern-.1667em\lower.7ex\hbox{E}\kern-.125emX}}

\title{An Architectural Style for Self-Adaptive Multi-Agent Systems}

\author{DANNY WEYNS\\KU Leuven, Belgium and Linnaeus University, Sweden\\
FLAVIO OQUENDO\\IRISA - Universit\'e de Bretagne-Sud, France}

\begin{abstract}

Modern distributed software systems often operate in dynamic environments in which operation conditions change continuously and subsystems may come and go at will, e.g. intelligent traffic management and multi-robot systems. To manage these dynamics, these systems have to self-adapt their structures and behaviors dynamically. While we have witnessed significant progress over the past decade in the manner in which such systems are designed, persistent challenges remain. In particular, dealing with distribution and decentralized control remains one of the major challenges in self-adaptive systems. This report presents an architecture style that supports software architects with designing architectures for a family of decentralized self-adaptive systems. The architecture style structures the software in a number of interacting autonomous entities (agents) that cooperatively realize the system tasks. Multi-agent systems derived from the architectural style realize flexibility (agents adapt their behavior and interactions to variable operating conditions) and openness (agents cope autonomously with other agents that enter and leave the system). The architectural style consists of five related patterns that distill domain-specific architectural knowledge derived from extensive experiences with developing various multi-agent systems. The architectural patterns are specified using $\pi$-ADL, a formal architectural description language supporting specification of dynamic architectures. This specification provides architects with a rigorous description of the architecture elements of the patterns, their interactions and behavior. We illustrate how we have applied the architectural style with excerpts of two cases from our practice: an experimental system for anticipatory traffic routing and an industrial logistic system for automated transportation in warehouse environments. 

\end{abstract}




\begin{document}

\sloppy
\setcounter{page}{1}

\begin{bottomstuff}
Authors' addresses: D.~Weyns, Department of Computer Science,
KU Leuven, 3000 Leuven Belgium and Linnaeus University, 351 95, Sweden;   danny.weyns@kuleuven.be. F.~Oquendo, IRISA Research Institute - Universit\'e de Bretagne-Sud, France, 
ENSIBS, BP 573, F-56017 Vannes Cedex, France; flavio.oquendo@irisa.fr.
\end{bottomstuff}
\maketitle

\section{Introduction}

Due to the increasing integration and dynamicity of software systems, providing and maintaining the required qualities of software constitutes a huge engineering challenge.  Modern distributed software systems, such as mobile multimedia applications, intelligent traffic management, and multi-robot systems usually comprise different subsystems that collaborate with each other as peers. Such systems typically have to operate in dynamic environments in which operation conditions change continuously  and subsystems may come and go at will. To manage these dynamics and maintain the required functionality and qualities, these systems have to self-adapt their structures and behaviors dynamically~\cite{1573858,Weyns2019}. Several researchers have argued that software architecture provides the right level of abstraction and generality to deal with the challenges of self-adaptability~\cite{Or,Rainbow,Kra:Mag,Weyns2012-1}. While we have witnessed significant progress over the past decade in the manner in which such systems are designed, persistent challenges remain~\cite{Dobson:2006,Huebscher:2008,sal:Tah:2009,1573856,deLemosetal:2012,Lemos2017}. In particular, dealing with distribution and decentralized control remains one of the major challenges in self-adaptive systems~\cite{Kra:Mag,Brun:2009,Weyetal:2012}. 

This report presents an architecture style that supports software architects with designing  a family of distributed self-adaptive systems in which control is decentralized. The architecture style distills the knowledge and expertise acquired from eight years of experience with designing and developing multi-agent systems. Multi-agent systems are a class of decentralized self-adaptive systems, comprising of autonomous agents that inhabit some complex dynamic environment, sense and act autonomously in this environment, and by doing so realize a set of goals or tasks for which they are designed~\cite{Mae95,Jen:Syc:Woo,Ster:Tav,1729509}. 
The architecture style structures the software in a number of such interacting autonomous agents that cooperatively realize the system tasks with two key quality properties: \textit{flexibility}, i.e. agents adapt their behavior and interactions to variable operating conditions, and \textit{openness}, i.e., agents cope autonomously with other agents that enter and leave the system. 

The architecture style consists of five related architectural patterns that provide generic solution schemes for recurring design problems. These architecture patterns make the application of the architecture style more operational. To document the architectural patterns we use a template inspired by the approach for documenting architecture styles presented in~\cite{Cle:Bac:Bas}. The elements of the architecture patterns and their relationships are specified in $\pi$-ADL~\cite{piADL}, a formally founded architecture description language. This specification provides a rigorous description of the architecture elements, their structures, and behaviors. The specification is type checked using the $\pi$-ADL.NET tool~\cite{piADLNET}.

The work presented in this report builds upon previous work~\cite{WeyWICSA2009} in which we have introduced a set of patterns for multi-agent systems. This report extends that work in multiple ways, including a refinement and elaborated discussion of the patterns of the architecture style, the application of the patterns to two applications including a system for anticipatory vehicle routing that was recently developed, and a systematic and complete specification of the structural and behavioral aspects of the architectural style in $\pi$-ADL. In addition, this report provides methodological guidelines for how to use the architecture style in practice, and a classification of related work on patterns for multi-agent systems.

\subsection*{Overview}

The report is structured as follows. In Section~\ref{sec:smulti-agent system} we introduce situated multi-agent systems, the family of decentralized self-adaptive systems that are supported by the architecture style. We discuss the background of the architecture style and introduce two applications that we have built with the style and use for demonstration in the remainder of the paper. Section~\ref{sec:structure} gives a high-level overview of architectural style, explains the template we use to document patterns, and introduces the necessary basics of $\pi$-ADL.  Section~\ref{sec:style} is the heart of the report, providing a detailed description of the five patterns of the architectural style. Section~\ref{sec:guidelines} provides methodological guidelines for using the architecture style. Section~\ref{sec:related} discusses related work, and Section~\ref{sec:conclusions} draws conclusions and looks at future work. For space reasons, we have omitted some parts of the $\pi$-ADL specification. For a complete $\pi$-ADL specification, we refer to the online Appendix A.

\section{Background}\label{sec:smulti-agent system}

This section introduces situated multi-agent systems, the family of self-adaptive systems supported by the architecture style. We give a brief overview of the applications that underly the architecture style and we introduce two applications from our practice that we use for illustration in the remainder of the report. Finally, we summarize the properties that determine the target domain of the architecture style.

\subsection{Situated Multi-Agent Systems}
A multi-agent system is in essence a system that is structured as a set of autonomous agents. \cite{Woo:Jen} define an agent as ``a computer system, situated in some environment, that is capable of flexible autonomous action in order to meet its design objectives.'' A multi-agent system is one that consists of a number of agents, which interact with one-another. \cite{Dur:Les} define a multi-agent system as ``a loosely coupled network of problem solvers (agents) that interact to solve problems that are beyond the individual capabilities or knowledge of each problem solver.'' Characteristics of multi-agent systems are as follows: (1) each agent has incomplete information or capabilities for solving the problem and, thus, has a limited viewpoint; (2) there is no system global control; (3) data is distributed; and (4) computation is asynchronous. Multi-agent systems are known for quality attributes such as adaptability, openness, robustness, and scalability, making them particularly interesting to deal with the demanding challenges of complex distributed software applications.

Situated multi-agent systems are one family of multi-agent systems. The focus of situated multi-agent systems is on modularization of agent behavior, efficient decision making, and indirect coordination~\cite{MaeGoals,Ferber99,545097,cofields,Par:Bru:Sau05,Wey09}. This contrasts with deliberative approaches of multi-agent systems that emphasize knowledge representation, rationality, planning, and direct communication~\cite{RGBDI,546581,Woo2000,Pok:Bra:Lam,Bor:2007}. A particularly interesting aspect of situated multi-agent systems is indirect coordination~\cite{Env,Wey:Omi:Ode,Weyns:2014:AEM:2951459.2951461}. Indirect coordination reduces the coupling between agents and provides a means for managing complexity by separation of concerns. Classical examples of indirect coordination are computational fields~\cite{189969,cofields} and digital pheromones~\cite{Bon:Dor:The,Par:Bru:Sau}. In the approach of computational fields, agents coordinate their behavior by following the shape of virtual force fields that are spread through a shared coordination infrastructure. Environment dynamics and movements of the agents induce changes in the surface of the fields, realizing a feedback cycle that influences the agents' movement. In the approach of digital pheromones, situated agents coordinate their behavior through digital markers (pheromones) in a shared coordination medium, similar to social ants. Agents can drop pheromones to form paths that can guide the agents to locations of interest. Digital pheromones evaporate over time. This mechanism ensures that paths to outdated information will fade away. These forms of indirect coordination are often constructed on top of basic mechanisms of shared memory based on tuple spaces~\cite{Carriero:1989,Schel2,Murphy:2006:LCM,MamZam2009}. 

\subsection{Underlying Expertise of the Architecture Style}

The architecture style embodies architectural knowledge gained from the design and development of various multi-agent system applications. We extensively used the Packet--World, a simple robotic application, as a study case for investigation and experimentation~\cite{PW}. We derived expertise from the design and development of a distributed peer-to-peer file sharing system~\cite{Sche2P2,WeyPer}. Our focus in this application was on coordination mechanisms inspired by principles of social ants. We have applied multi-agent systems in several experimental robotic applications focussing on the roles agents play to set up collaborations~\cite{Stee,WeySELMAS04}. We have employed a decentralized multi-agent system architecture in an industrial transportation system for controlling automatic guided vehicles~\cite{Wey05,2008OOPSLA}. A particular interest in this industrial application was on advanced coordination mechanisms that enable situated agents to flexibly adapt their behavior to changes in the stream of tasks that enter the system and other environmental dynamics. We have been using agents in the domain of intelligent transportation systems, for example for traffic jam monitoring~\cite{MACODO-arch}, supply chains~\cite{Haesevoets1}, and anticipatory vehicle routing~\cite{1538945}. The focus here is on situated agents that exploit a coordination medium to dynamically organize themselves based on the changing context in which they are situated.  Recently, we have also been studying how we can apply similar ideas in the management of dynamic supply chains~\cite{Hae12}.

We now introduce two cases that we use to demonstrate how we have applied the architectural style and realized the patterns in our work.  We start by introducing an experimental system for anticipatory vehicle routing. Then we introduce an industrial logistic system for automated transportation.

\subsubsection{Anticipatory Vehicle Routing System}

\begin{figure}[ht!]
\centering
\resizebox{0.92\textwidth}{!}
{\includegraphics{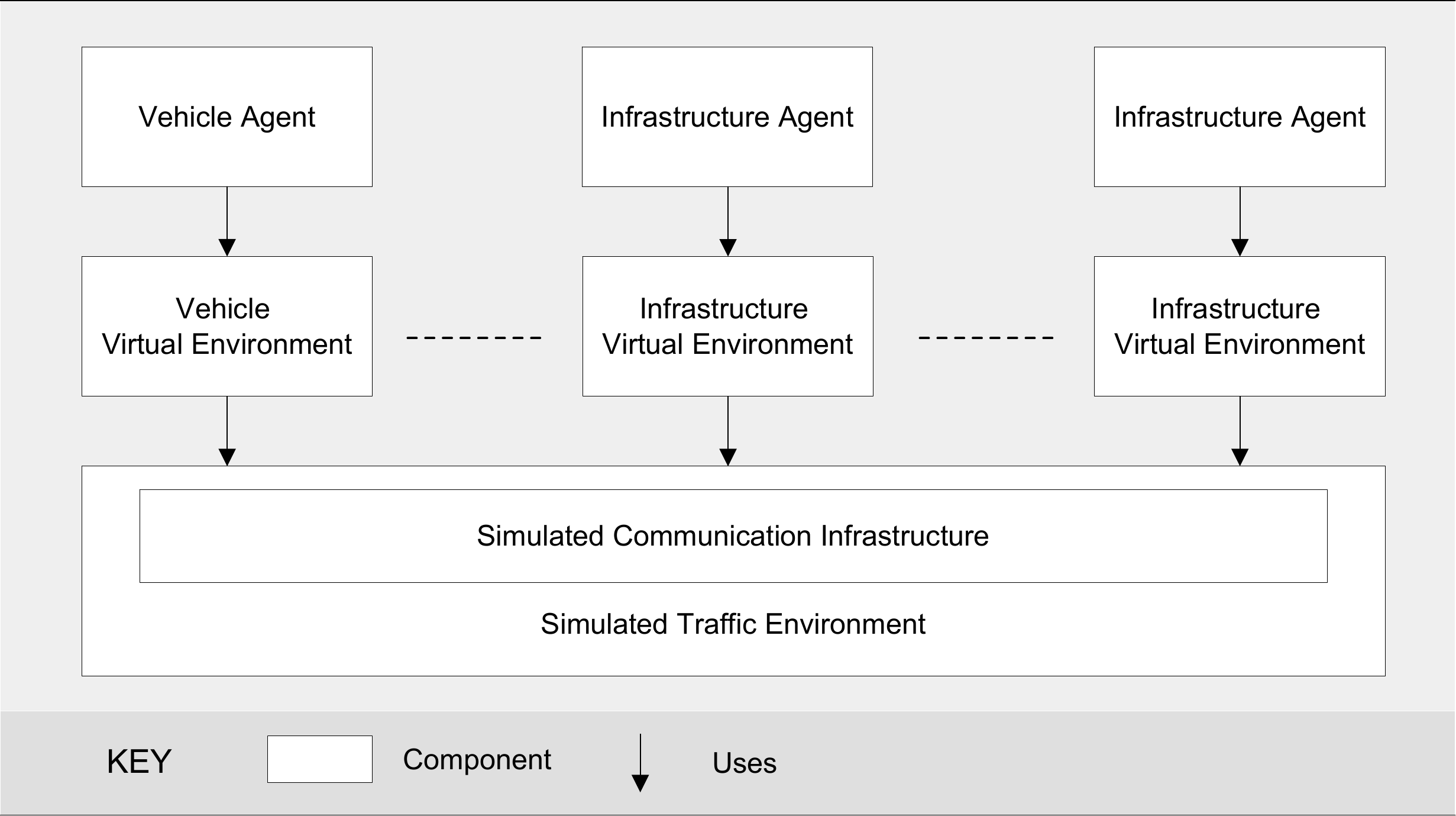}}
\caption{Component model of the simulated anticipatory vehicle routing system}\label{fig:graph}
\end{figure}

The first application is an experimental system to avoid traffic congestion based on anticipatory vehicle routing~\cite{KULeuven-167666,1538945}. %
Fig.~\ref{fig:graph} shows a high-level component model of the system.

This application is situated in the domain of intelligent transportation systems, a worldwide initiative to exploit information and communication technology to improve traffic.\footnote{http://www.itsworldcongress.com/} 
The large number of vehicles today, the limited capacity of the road networks, and unpredictable dynamics of the traffic context (such as a fluctuating amount of traffic, changing behavior of drivers, traffic jams, and road accidents) make traffic routing a particularly challenging problem. Existing technology such as satellite navigation devices exploit broadcast traffic information to trigger vehicles to reroute and bypass traffic jams. Although such mechanisms allow substantial performance gain already, they only react upon traffic jams after they occurred. In our research, we study anticipatory vehicle routing that aims to encompass this by using forecast of traffic density.

The approach presented here assumes that the road infrastructure is equipped with electronic devices which provide some computation power and are connected through a network. Vehicles are equipped with smart devices which can sense that vehicle's local traffic conditions and which can communicate with other vehicles and electronic devices of the road infrastructure.

On each vehicle device, a vehicle agent is deployed (see Figure~\ref{fig:graph}) that has knowledge about that vehicle's start location and destination, and its current state such as position and speed. The vehicle agent provides information to the driver to reach its destination with minimal congestion overhead. On each infrastructure device, an infrastructure agent is deployed that has knowledge about the traffic conditions of a part of the road network and maintains bookings for future traffic.

Each vehicle agent explores the relevant paths in the environment, and based on the evaluation of the possible alternatives, it chooses one path and records it as an intention. Therefore, the vehicle agent informs the corresponding infrastructure agents of the time the vehicle intends to occupy the corresponding road elements, i.e.~the agents need to book these reservations. To explore paths and book reservations vehicle agents use so called ant agents. Ant agents are smart messages that travel on behalf of a vehicle agent along a path in the environment visiting a sequence of infrastructure agents~\cite{Holvoet:2009}. 

A vehicle virtual environment reflects the real traffic environment. This virtual environment is a distributed software entity that is deployed on each device in the traffic system. Vehicle agents and infrastructure agents use the virtual environment to enable  the ant agents to explore feasible paths in the environment and make bookings for the selected paths. 

Flexibility and openness are key requirements in this domain as vehicles face continuously changing operating conditions, such as changing traffic conditions, new vehicles that enter and leave the system, etc. Therefore, vehicle agents may revise there intentions and book alternative routes when needed. To that end, the vehicle agents 
send new ant agents at regular times to explore changes in the environment and refresh bookings.  This mechanism makes the system robust to dynamics in the traffic environment.

Our experiments are done in simulation of a real-world traffic environment, which include a model of the physical traffic environment and the communication network. The simulated environment maintains the state of elements of the physical system, provides the means for monitoring the state of elements in the system, determines the outcome of the actuator actions of the agents, and it collects and delivers messages.

\subsubsection{Automated Transportation System}

The second application is an industrial transportation system that was developed between 2004 and 2007 in a collaboration between researchers of DistriNet Labs and a team of engineers and developers of Egemin\footnote{http://www.egemin.com/}, a leading company that provides full life cycle support for automated transportation systems~\cite{Wey05,2008OOPSLA,Wey09}.

An automated transportation system consists of a number of automatic guided vehicles (AGVs) that need to work together to transport loads in an industrial environment. Transport tasks are generated by client systems, typically an enterprise resource planning (ERP) system. The main functionalities that an AGV transportation system has to fulfill are assigning incoming transportation tasks to appropriate AGVs, routing the AGVs through the warehouse efficiently while avoiding collisions and deadlocks, and recharging the AGVs' batteries.

An AGV transportation system has to deal with dynamic and changing operating
conditions. The stream of transportation tasks that enters the system is typically irregular and unpredictable, AGVs can leave and re-enter the system for maintenance, production machines may have variable waiting times, etc. All kinds of disturbances can occur, supply of goods can be delayed, certain areas in the warehouse may temporarily be closed for maintenance services, loads can block paths, AGVs can fail, etc. Despite these challenging operating conditions, the system is expected to operate efficiently and robustly.


Traditionally, the AGV systems deployed by Egemin are directly controlled by a central server. The server plans the schedule for the system as a whole, dispatches commands to the AGVs, and continually polls their status. System operators monitor the system and intervene whenever needed.  This results in reliable solutions. The central point of control also enables easy diagnosis of errors. However, a shift in user requirements challenges the centralized architecture. Customers increasingly request for systems that are able to adapt their behavior with changing circumstances autonomously, i.e. without intervention of an operator. For example, the system should adapt itself when the stream of transportation tasks changes dynamically, or when disturbances occur.  Such self-adaptability requires flexibility and openness of the solution. Flexibility enables the system to deal with dynamic operating conditions autonomously, and openness enables it to deal with AGVs entering and leaving the system.

\begin{figure}[t!]
\centering
\resizebox{0.92\textwidth}{!}
{\includegraphics{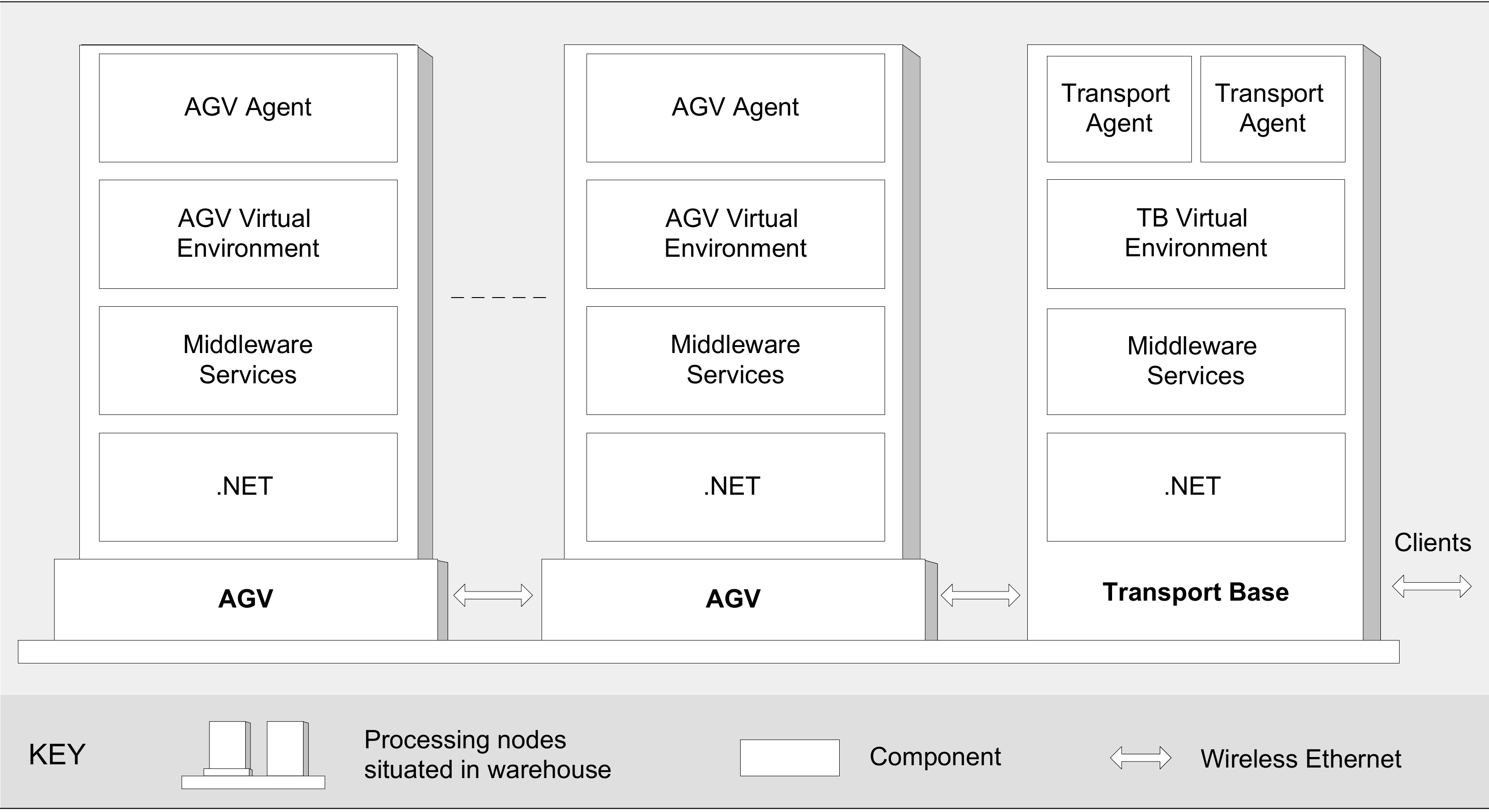}}
\caption{High-level deployment model of a transportation system} \label{fig:sa-model}
\end{figure}

Applying a situated multi-agent system opens perspectives to improve flexibility and openness of the system: the AGVs can adapt themselves to the current situation in their vicinity, order assignment is dynamic, the system can deal autonomously with AGVs leaving and reentering the system, etc.
Fig.~\ref{fig:sa-model} shows a high-level deployment model of the agent-based system.

Each vehicle is controlled by an AGV agent, and each transportation task in the system is represented by a transport agent that is deployed at the transport base. To realize the system requirements, AGV agents and transport agents have to coordinate, e.g.~for task assignment, for collision avoidance, etc. Therefore, the agents exploit a local virtual environment. The states of neighboring local virtual environments are synchronized using middleware services that run on top of the .NET platform.

\subsection{Target Domain of the Architecture Style}\label{TargetDomain}

We summarize the key characteristics and requirements shared by the systems that underly the architecture style: 
\begin{itemize}
\item The software systems are subject to dynamic and changing operating conditions, such as dynamically changing workloads and variations in availability of resources and services.
\item Activity in the systems is inherently localized, i.e.~global control or access to resources is difficult to achieve or even infeasible.
\item Important stakeholder requirements are (1) flexibility: the software systems have to adapt to variable operating conditions autonomously; and (2) openness: the software systems have to cope autonomously with parts that come and go during execution.
\end{itemize}
In addition, we consider the following constraints in terms of the domain: 
\begin{itemize}
\item The speed of changes in the environment are orders of magnitude lower than the speed of communication
and the execution of the control software.
\item Quasi continual communication access to the distributed software system is available, e.g., via a LAN.
\item We consider distributed software systems within a single ownership domain. 
\end{itemize}
These characteristics and constraints determine the target domain (family of systems) that are supported by the architecture style for multi-agent systems.


\section{Overview of the Architecture Style}\label{sec:structure}

In this section, we give a general introduction of the architectural style. We start with a brief overview of  the patterns that constitute the architecture style. Then, we outline the pattern template that we use for the detailed description of the patterns. We conclude the section with a brief overview of the basic language primitives of $\pi$-ADL.

\subsection{Patterns of the Architecture Style}
\begin{figure}[ht!]
\centering
\resizebox{.74\textwidth}{!}
{\includegraphics{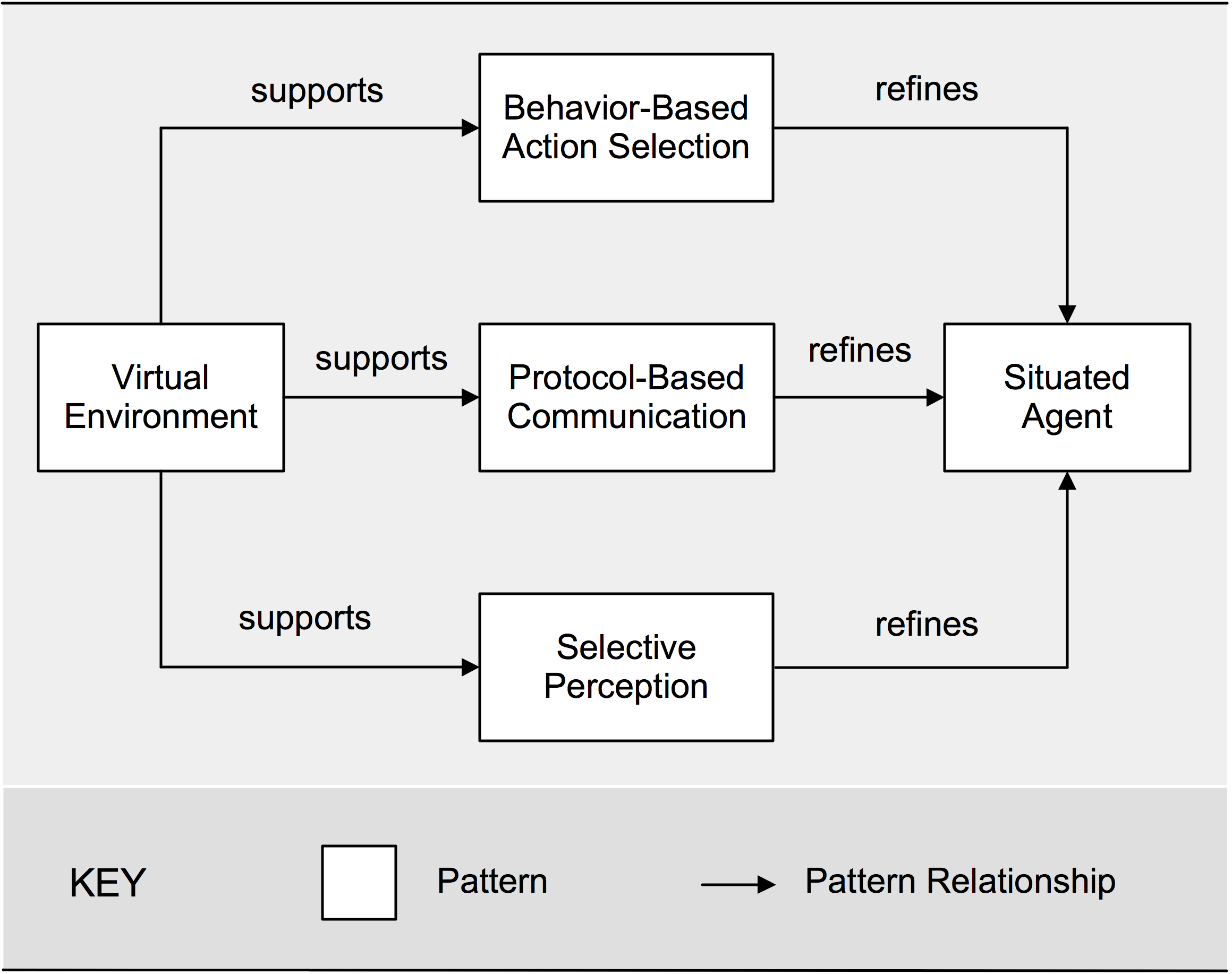}}
\caption{Overview of the patterns of the architectural style}
\label{fig:style-overview}
\end{figure}

Fig.~\ref{fig:style-overview} shows an overview of the different patterns of the architecture style with the relationships between the patterns.

The basic patterns of the architecture style are \emph{Situated Agent} and \emph{Virtual Environment}. A situated agent is an autonomous problem solving entity that encapsulates its state and controls its behavior. The responsibility of an agent is to realize the application specific tasks it is assigned. Situated agents are cooperative entities that are able to flexibly adapt their behavior and interactions with changing conditions. Agents are situated in a virtual environment that maintains a virtualization of the relevant parts of the world and serves as a coordination medium for the agents, i.e.~the virtual environment mediates both the interactions among agents and the access to resources.

Besides the basic patterns, there are three additional patterns that refine the situated agent pattern: \emph{Selective Perception}, \emph{Protocol-based Communication}, and \emph{Behavior-Based Action Selection}. A situated agent can use selective perception to sense the  environment and update its knowledge about the world. 
Situated agents can use protocol-based communication to exchange messages according to prescribed communication protocols, i.e.~well-defined sequences of messages. 
Finally, a situated agent can use behavior-based action selection to decide about the actions it wants to perform in the environment. 
Selective perception, protocol-based communication, and behavior-based action selection are supported by the virtual environment. 

The patterns have several variability points. For example, for the design of agents that only communicate indirectly via the virtual environment, the protocol-based communication pattern can be omitted. We elaborate on variability in Section~\ref{sec:guidelines}.

\subsection{Pattern Template}

To document the patterns of the architecture style we use the following template:
\begin{enumerate}
    \item The name of the pattern.
    \item A primary presentation that shows the elements and their relationships in the pattern. We use component and connector models to describe the pattern's units of execution.
    \item A description of the architectural elements and their responsibilities. We illustrate the description with examples from the two cases introduced in Section~\ref{sec:smulti-agent system}. The architectural elements are specified in $\pi$-ADL. Some fragments are omitted for brevity. For a complete formal specification, we refer to the online Appendix A. 

    \item A rationale that motivates the design of the pattern.
\end{enumerate}
This pattern template follows the approach for documenting architectural styles presented in~\cite{Cle:Bac:Bas}.
We discuss related patterns from the field of multi-agent systems in Section~\ref{sec:related}. 

\subsection{$\pi$-ADL in a Nutshell}

The patterns of the architectural style are specified in $\pi$-ADL, a formal architectural description language. It is a theoretically well-founded language based on the higher-order typed $\pi$-calculus, supported by a specification and verification toolset providing services for automated checking of architectural properties.

Formal specification improves the understanding of the specification, enables reasoning about the patterns, and provides a firm foundation for employing the patterns in practice. The patterns are type checked using the $\pi$-ADL.NET tool~\cite{piADLNET}. In this section, we briefly explain the main language constructs of $\pi$-ADL. We limit the explanation to the subset of language constructs that we use to describe the patterns.
\begin{itemize}
\item[$\bullet$] \emph{abstraction}: the basic architectural element in $\pi$-ADL. Abstraction is a unit of execution.

\item[$\bullet$] \emph{component}: the basic computational element that we use in the description. A component is an architectural abstraction that delivers some system functionalities.

\item[$\bullet$] \emph{connection}: a basic interaction point. A connection provides a communication channel between architectural elements that allows value passing. Values are sent and received through connections via the out and in prefixes respectively. Value passing via a connection is synchronous.

\item[$\bullet$] \emph{port}: an interface between a component and its environment. Ports are described in terms of connections. Protocols may be enforced by ports and among ports.

\item[$\bullet$] \emph{unifies}: expresses the linking of two connections. Unifies connects connections together.

\item[$\bullet$] \emph{relays}: expresses the binding between external ports and ports of components. If connections have the same names in different ports, identifying ports is enough to express unifications.

\item[$\bullet$] \emph{type}: a value type which can be declared either a basic type or a constructed type. A type is declared within the scope of a behavior.

\item[$\bullet$] \emph{Integer, Boolean, and String}: regular basic types.

\item[$\bullet$] \emph{Any}: an unspecified data type, represents any arbitrary combination of data types. We use any as an abstract data type that has to be made concrete when the patterns are applied to a concrete domain.

\item[$\bullet$] \emph{view}: a collection of named elements, possibly of different types.

\item[$\bullet$] \emph{sequence}: an indexed collection of elements of the same type.

\item[$\bullet$] \emph{set}: an abstract collection of elements of a particular type.

\item[$\bullet$] \emph{location}: an abstract storage of elements of a particular type.

\item[$\bullet$] \emph{behavior}: an expression of the process of a component in terms of actions and interactions with its environment through connections.

\item[$\bullet$] \emph{function}: a part of functionality within a behavior which performs a specific task. A function may take input of different types and produce output of a particular type.

\item[$\bullet$] \emph{compose}: expresses the capability of a behavior to parallel compose the capabilities of multiple behaviors. The composed behaviors can proceed independently and can interact via attached connections.

\item[$\bullet$] \emph{choose}: expresses the capability of a behavior to choose the capability of one of set of behaviors. If all the sub-behaviors are blocking on an input, the first one to resume execution will continue while the others will terminate.

\item[$\bullet$] \emph{unobservable}: expresses the capability to enact an action invisibly, i.e.~internally, silently. Non-observable behavior is internal to an abstraction.

\end{itemize}



\section{Architectural Style}\label{sec:style}

We now present the patterns of the architectural style using the pattern template. We start with the two basic patterns: Virtual Environment and Situated Agent. Then we explain the three additional patterns that refine a situated agent: Selective Perception, Protocol-Based Communication, and Behavior-Based Action Selection.

\subsection{Virtual Environment}

\subsubsection{Primary Presentation}
\begin{figure}[ht!]
\centering
\resizebox{.65\textwidth}{!}
{\includegraphics{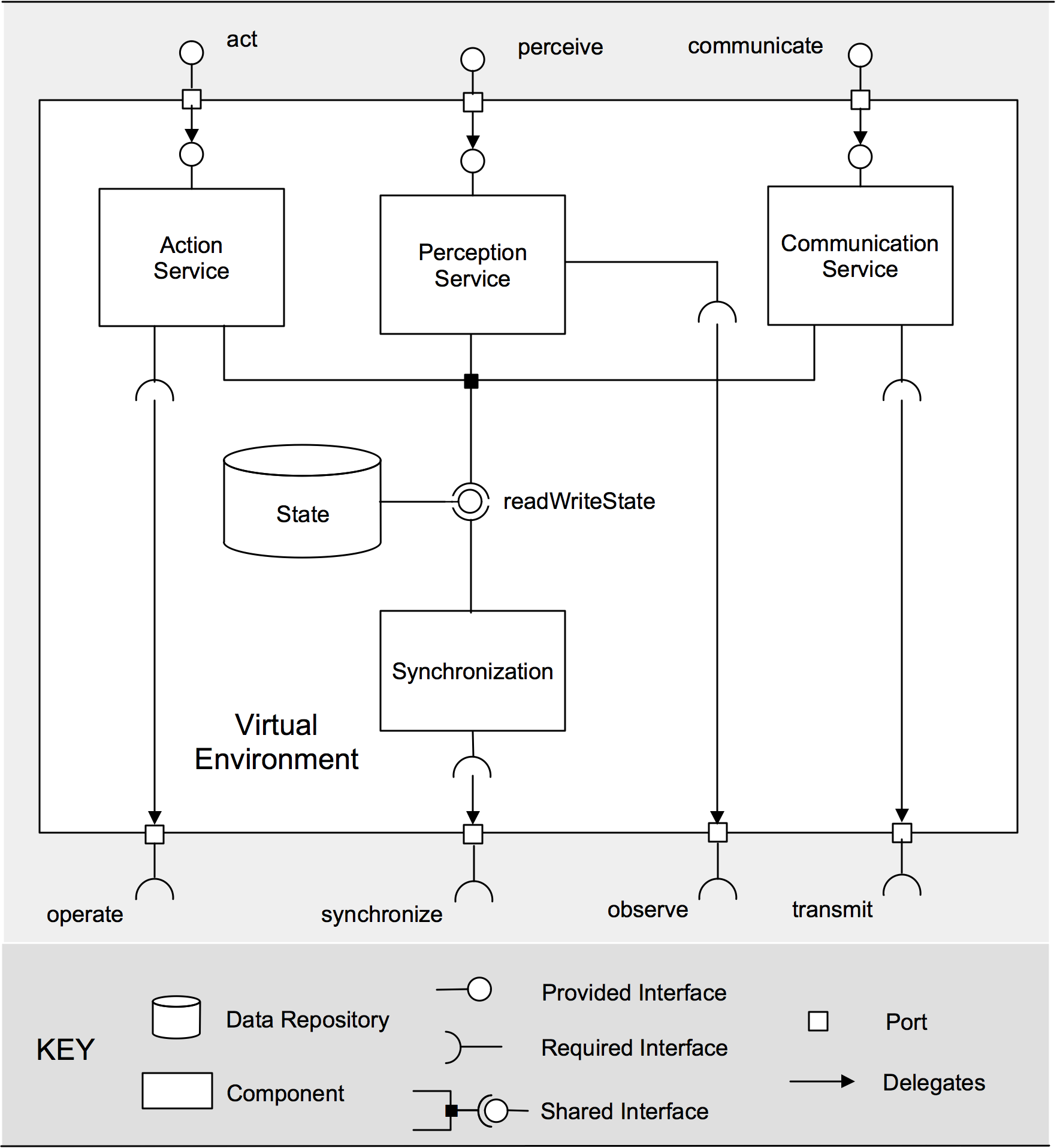}}
\caption{Primary presentation of virtual environment}
\label{fig:virtual-environment}
\end{figure}

\mbox{\ }\vspace{6pt}\newline
The primary presentation of the virtual environment pattern is shown in Figure~\ref{fig:virtual-environment}.

The virtual environment provides a domain-specific infrastructure that supports the activities of situated agents (perception, communication, performing actions). In particular, it provides a virtualization of the  external environment, possibly augmented with support for coordination.\footnote{\textit{Environment} is an overloaded term in software engineering. The term is used for development environment, execution environment, physical environment, etc. In this report, we refer with \textit{environment} (or \textit{external environment}) to the part of the external world (which may comprise both physical or software elements) with which the system interacts, and in which the effects of the system will be observed and evaluated~\cite{Jackson:1997}.} For example, the virtual environment may represent the physical context as a map and this map may be equipped with facilities for agents to add marks on particular locations, supporting indirect coordination among the agents. The virtual environment requires middleware services for observing the external environment, transmitting messages, and executing actions in the environment. At the top level, the virtual environment is defined as follows:

\footnotesize{
\begin{verbatim}
component VirtualEnvironment is abstraction ()
{
     type Focus is view[focusName : String, focusParams : Any].
     type Sense is view[agentId : String, focus : Focus].
     type Representation is Any.
     type Action is view[agentId : String, actionName : String,
            actionParams : Any].
     type Operation is Any.
     type Message is view[id : Integer, sender : String, receiver : String,
            protocol : String, performative : String, content : Any].
     type Observation is Any.
     type Observed is Any.
     type StateItem is view[name : String, val : Any].
     type StateItems is set(StateItem). 
     type SynchronizationUpdate is Any.
     type Transmission is Any.

     port perceive is { connection sense is in(Sense).
                        connection sensed is out(Representation). }
     port act is { connection action is in(Action) }.
     port communicate is { connection sendMsg is in(Message).
                           connection receiveMsg is out(Message) }.
     port operate is { connection operation is out(Operation) }.
     port synchronize is { connection syncOut is out(SynchronizationUpdate).
                           connection syncIn is in(SynchronizationUpdate) }.
     port observe is { connection observation is out(Observation).
                       connection observed is in(Observed) }.
     port transmit is { connection transmission is out(Transmission).
                        connection deliver is in(Transmission) }.

     //composition
     behavior is compose {
               actionService is ActionService()
         and   perceptionservice is PerceptionService()
         and   communicationservice is CommunicationService()
         and   state is State()
         and   synchronization is Synchronization()
     }  where  {
               sense relays perceptionService::sense
         and   perceptionService::readWriteState unifies state::readWriteState
         and   perceptionService::sensed relays sensed
         ...
        }
     }
}
\end{verbatim}
}
\normalsize
The $\pi$-ADL specification first defines the different types and ports. Some types include the \textit{Any} data type. These types have to be made concrete when the pattern is applied to a concrete domain. Then, the composition of the various component instances is defined (for brevity, only the composition of the perception service is showed). The components are further refined and explained in the following subsection.

\subsubsection{Architectural Elements}\label{sec:arch-elems-ve}
%
%
\mbox{\ }\vspace{6pt}\newline
The \textbf{State} repository has a central role in the virtual environment. The repository contains data that is shared between the components of the virtual environment. The $\pi$-ADL specification of state is defined as follows:

\footnotesize{
\begin{verbatim}
component State is abstraction()
{
   type StateItem is view[name : String, val : Any].
   type StateItems is set(StateItem).

   port readWriteState is {
          connection stateTemplate is in(StateItems).
          connection readState is out(StateItems).
          connection writeState is in(StateItems) }.
          assuming { protocol is (via stateTemplate receive StateItems.
                                    true*.via readState send StateItems)* 
                                    or (via writeState receive StateItems)* }.

   stateRepository is location(set(StateItems)).

   behavior is {
      write is function( state : StateItems,
        repository : set(StateItems) ) :
           set(StateItems) { unobservable }.
      read is function( template : StateItems,
         repository : set(StateItems) ) :
           StateItems { unobservable }.
      choose
      {
        //read state
        via stateTemplate receive state_template : StateItems.
        via readState send read(state_template, stateRepository).
        behavior().
      or
        //write state
        via writeState receive state_items : StateItems.
        stateRepository := write(state_items, stateRepository).
        behavior().
      }
   }
}
\end{verbatim}
}
\normalsize
State items are abstract data that represent elements of interest maintained by the virtual environment. State items are stored in the state repository. The state repository provides a port that allows clients to read state based on a state template, and write state. The behavior specifies how reading and writing is executed. Reading requires a template of state items to select the state items of interest. The internal behavior of read and write are invisible and have to be made concrete for a particular domain.

State stored in the state repository typically includes an abstract representation of external resources and additional state that is used for coordination purposes. An example of state related to external resources in the vehicle routing system is a map of the traffic environment.  The physical road network is mapped onto a graph representation of discrete road elements, i.e.~road segments and crossroads, that are connected via edges. The road network is concretely realized as follows: 

\footnotesize{
\begin{verbatim}
  public class Edge<RoadElement> {
     private int id;
     private RoadElement elem;
   
     ...
  }
 
  public class RoadNetwork<RoadElement> {
     private ArrayList<Edge<RoadElement>> edges;
     public RoadNetwork(){
        edges = new ArrayList<Edge<RoadElement>>();    
     }
     ...
  }
\end{verbatim}
}
\normalsize

\begin{figure}[!t]
	\centerline{
		\subfloat[Rendering of the OpenStreetMap data]{
			\includegraphics[width=2.7in,height=2.3in]{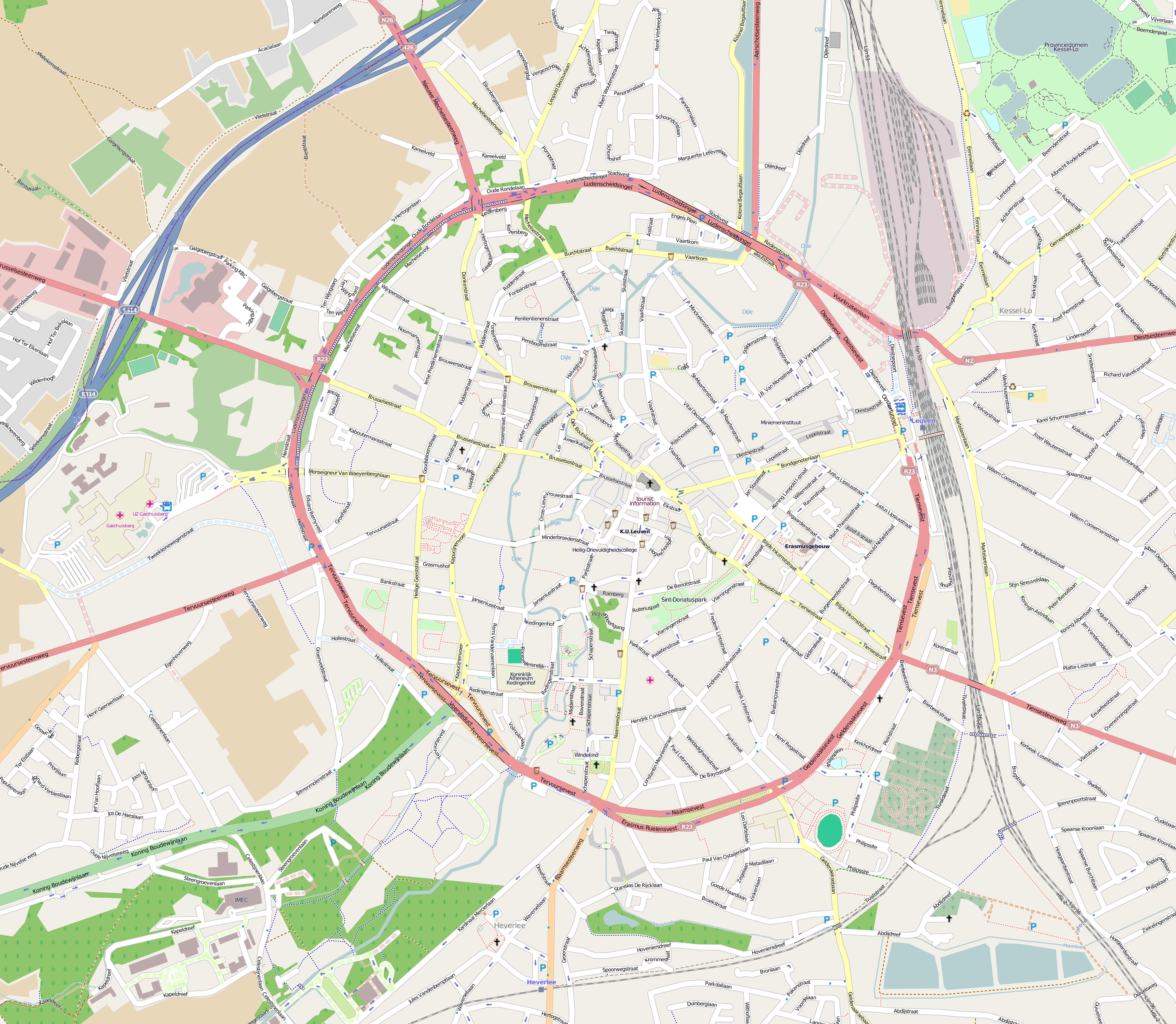}%
			\label{fig:map:leuven:rendered}
		}
		\hfil
		\subfloat[Graph model based on OpenStreetMap data]{
			\includegraphics[width=2.7in,height=2.3in]{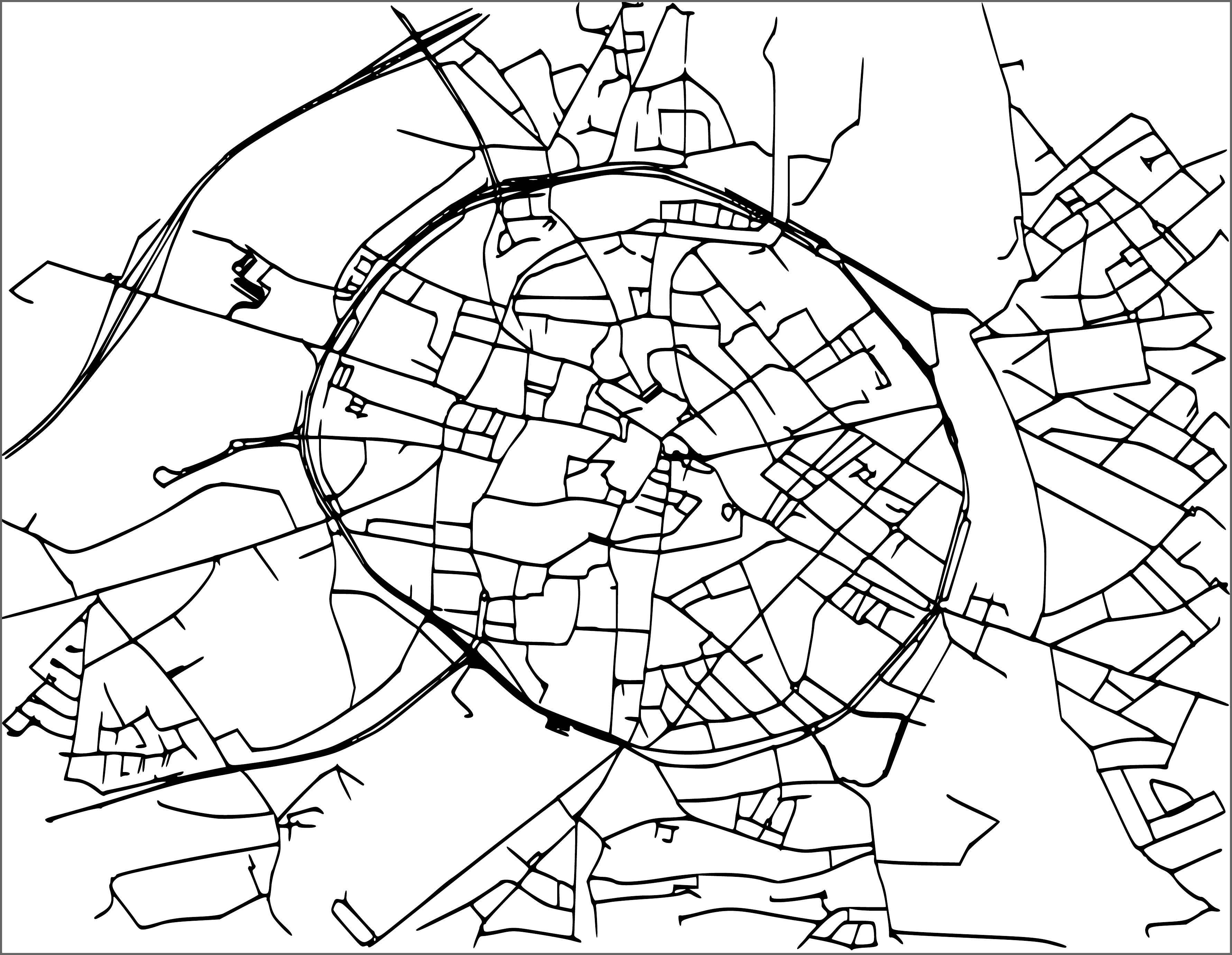}%
			\label{fig:map:leuven:graph}
		}
	}
	\caption{Example of a graph model used in the vehicle routing system.  
		The figure on the left shows a rendering of the OpenStreetMap data of the city center of Leuven.  
		The figure on the right shows the graph with over 1600 roads.}
	\label{fig:map}
\end{figure}

Edge has an id and a road element of the type RoadElement, corresponding to the name and value of a state item. The road network models a state repository that comprises a list of edges. Road network implements regular accessors and mutators to read and write state. For our case study, we used OpenStreepMap to generate the graph of the city of Leuven, which consists of over 1600 road elements as shown in Figure~\ref{fig:map}.
%
In the vehicle routing system, additional state maintained by the state repository are the current position of the vehicle on the map, and road segments that are marked indicating that they are temporarily blocked.
%

\begin{figure}[h!]
\centering
\resizebox{.7\textwidth}{!}
{\includegraphics{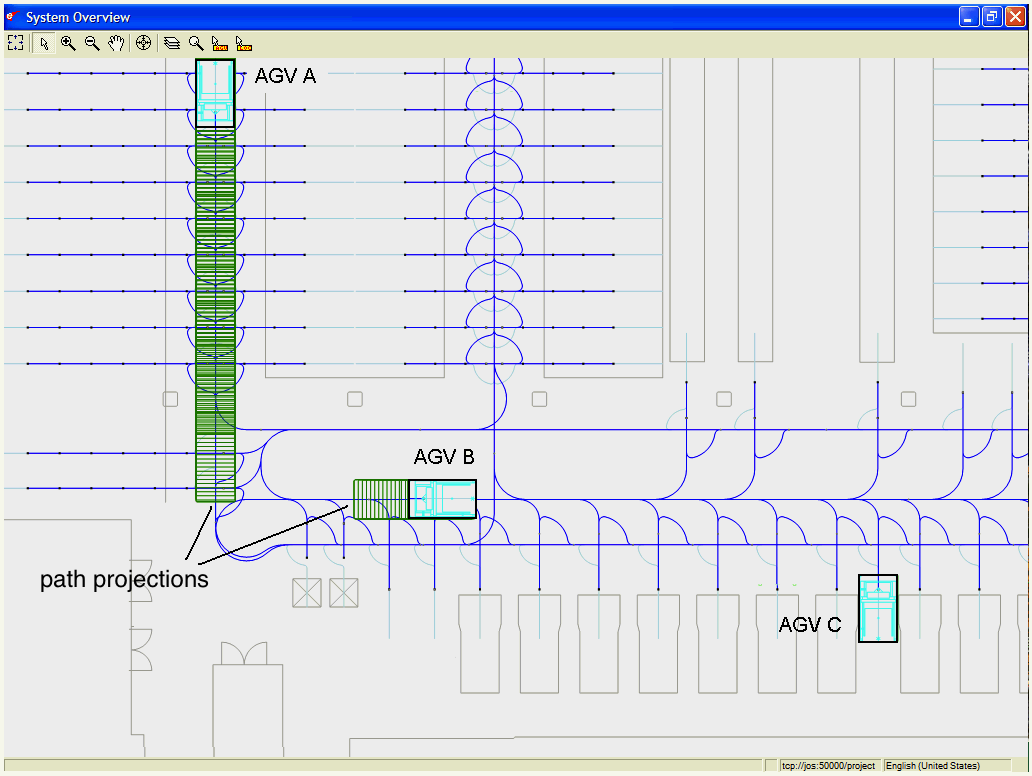}}
\caption{Path projections in the virtual environment.}
\label{fig:hulls}
\end{figure}

Similarly in the AGV transportation system, the state related to external resources includes a map of the warehouse that is modeled as a graph of segments. A segment represents an atomic part of the warehouse infrastructure along which an AGV can be instructed to drive (typically 3 to 5 meter). Examples of additional state are virtual marks situated on the map. As an example, consider Fig.~\ref{fig:hulls} that shows a fusion view\footnote{The view is generated on a remote  machine that collects the state of the virtual environment of the AGVs via a wireless network and fuses the information into one image.} of the AGV local virtual environments of three AGVs.  

To avoid collisions, AGV agents have to lock the segments they want to drive. Therefore, the virtual environment uses path projections that are realized as  follows: 

\footnotesize{
\begin{verbatim}
  public class PathProjection {
     private int id;
     private int priority;
     private Hull hull;
     private ArrayList projection;
     private Status status;     
     ...
     public void lock()  { ... } 
     public void clear(ArrayList segments)  { ... } 
     ...
  }
\end{verbatim}
}
\normalsize

A path projection has an identifier, a priority that is used to give precedence to AGVs with important transportation tasks, a hull that represents the physical area an AGV occupies on the floor, a projection that consists of an array of segments of the intended path, and finally a status which is \emph{requested} or \emph{locked}. When an AGV agent projects the path projection in the AGV local virtual environment, the path projection will have status requested. The local virtual environment then coordinates with other local virtual environment and marks the path projection \emph{locked} when the conditions for the AGV are safe to move on. This coordination is realized by the synchronization component discussed below. When the AGV drives forward, the path projection is gradually cleared. 
\mbox{\ }\vspace{6pt}\newline 
The \textbf{Perception Service} provides the functionality for agents to sense the environment. The $\pi$-ADL specification of the perception service is defined as follows:

\footnotesize{
\begin{verbatim}
component PerceptionService is abstraction ()
{
     type Focus is view[focusName : String, focusParams : Any].
     type Sense is view[agentId : String, focus : Focus].
     type Representation is Any.
     type Observation is Any. 
     type Observed is Any.
     ... 
          
     port perceive is { connection sense is in(Sense).
                        connection sensed is out(Representation). }
     port observe is { connection observation is out(Observation).
                       connection observed is in(Observed) }.
     ...
    
     perceptionTypes is location(set(StateItems)). 

     behavior is {       
         analyzePerception is function ( sense : Sense ) : 
           StateItems { unobservable }.	  
         virtualPerception is function ( items : StateItems, 
           types :  set(StateItems) :
             Boolean { unobservable }.
         generateRepresentation is function( sense : Sense, 
           state : StateItems) :
             Representation { unobservable }.
         ...

         via sense receive sense_request : Sense.
         sense_items := analyzePerception( sense_request ). 	
         if virtualPerception( sense_items, perceptionTypes )
             then  {
                  //perceive local state repository
                  unobservable.
                  via stateTemplate send sense_items.
                  via readState receive perceived_state.
                  via sensed send generateRepresentation(
                        sense_request, perceived_state ).
                  behavior().
             }
             else {
                  //perceive external environment
                  unobservable.
                  via observation send
                        generateObservation( sense_items ).
                  unobservable.
                  via observed receive external_observed.
                  observation_state :=
                    analyzeObserved( external_observed ).
                  via sensed send
                    generateRepresentation( sense_request, 
                      observation_state ).
                  behavior().
             }
  	}
}
\end{verbatim}
}
\normalsize
The perception service supports \emph{selective perception}. Selective perception enables an agent to direct its perception at the relevant aspects according to its current task. This facilitates better situation awareness. To direct its perception an agent selects a \emph{focus}. A focus has a set of domains-specific parameters that determine aspects such as element types of interest, operating range, etc. Sensing results in a representation. A representation is a domain-specific data structure that represents the sensed elements in a form that can be interpreted by the agents. 

When an agent invokes a perception request, the perception service analyzes it and determines whether it concerns a virtual perception of the local state or a perception of the external environment. In case of a virtual perception, a representation is generated based on state derived from the repository. In case of an external perception, an observation is generated, and a representation is generated from the observed state. An observation is a domain-specific command that allows observing state of the external environment. 

Examples of foci in the AGV transportation system are a focus to observe the positions of nearby vehicles and a focus to observe the status of an operating space. An example in the vehicle routing system is a focus to sense a path on the road infrastructure:   

\footnotesize{
\begin{verbatim}
    public PathFocus(String name, int maxDist, Node from, Node to) { 
       ... 
    }
    pfocus = new PathFocus("paths", 1500, n15, n125); 
\end{verbatim}
}
\normalsize

This focus selects the traffic routes with a maximum distance of 1500 m from a given node (n15) to another node (n125). The nodes refer to the elements of the road elements of the road network. Nodes are an abstraction of road elements used by the agents. The perception service maps nodes to road elements of the road network maintained by the state repository of the virtual environment. Note that the result of a perception with this focus may dynamically change when certain road areas are temporary not accessible as reflected in the map. Such map updates can be maintained by the synchronization components (discussed below).  

The representation that results from sensing a path on the road infrastructure is an object of graph paths that is realized as follows:     

\footnotesize{
\begin{verbatim}
    public class GraphPath {
       private ArrayList<Node> path; 
       ...
    }
    public class GraphPaths {
       private ArrayList<GraphPath> paths; 
       ...
    }
\end{verbatim}
}
\normalsize

Graph paths contains a list of paths, each path consisting of a list of nodes. Agents can interpret graph paths to support their decision making. 
%

Both for the AGV transportation systems and the vehicle routing system, we only use virtual perceptions.  Synchronization of the state of the virtual environment with the external environment is realized by the synchronization component (discussed below). An example of a virtual perception in the AGV transportation systems is a perception request to check the status of a path projection (see Fig.~\ref{fig:hulls}). An example in the vehicle routing system is a perception request to sense traffic roads with a path focus:

\footnotesize{
\begin{verbatim}
    public class PathSense( PathFocus pf ) { 
        ...
    }
    psense = nes PathSense(pfocus); 
    virtualEnv.sense(id, psense); 
\end{verbatim}
}
\normalsize
%
\mbox{}
\vspace{5pt}\newline
\textbf{Action Service} is responsible to deal with agents' actions. The $\pi$-ADL specification of the action service is defined as follows:

\footnotesize{
\begin{verbatim}
component ActionService is abstraction ()
{
     type Action is view[agentId : String, actionName : String,
            actionParams : Any].
     type Operation is Any.
     ...

     port act is { connection action is in(Action) }.
     port operate is { connection operation is out(Operation) }.
     ...

     actionTypes is location(set(StateItems)). 

     behavior is {
          analyzeAction is function ( act : Action ) :
            StateItems { unobservable }.
          virtualAction is function( state : StateItems,
            types :location(set(StateItems)) :
              Boolean { unobservable }.
          updateState is function( state: StateItems ) :
            StateItems { unobservable }.
          generateOperation is function( state : StateItems ) :
            Operation { unobservable }.
         
         action_items : StateItems.

         via act receive action_request : Action.
         action_items := analyzeAction( action ). 	
         if virtualAction( action_items, actionTypes )
             then {
                  //apply action to local state
                  unobservable.
                  via writeState send updateState( action_items ).
                  behavior().
             }
             else {
                  //apply external action
                  unobservable.
                  via operate send generateOperation( action_items ).
                  behavior().
             }
     }
}
\end{verbatim}
}
\normalsize

Situated agents perform actions to realize their tasks. When an agent invokes an action, the action service analyzes the action based on the action types to determine whether the action is concerned with modifying the state of the virtual environment (virtual action) or modifying the state of external resources (external action). In the former case, the local action is executed on the state repository of the virtual environment. In the latter case, an operation is generated and executed. An operation is an action that can be executed in the external environment. 

An example of a virtual action in the AGV transportation system  is an agent that projects a requested path on the map in the AGV local virtual environment (see Fig.~\ref{fig:hulls}). 

\footnotesize{
\begin{verbatim}
    public Project(String name, int prio, Hull hull, Path path) { 
       ... 
    }
    ...
    project = new Project("project", 0, hull, path);  
    virtualEnv.invoke(id, project);  
\end{verbatim}
}
\normalsize

The project action has a name, the priority of the transport currently performed by the requesting AGV, the hull of the AGV, and the path for the projection (i.e., a list of nodes that correspond to segments). The excerpt illustrates how a concrete action is instantiated and then executed on the virtual environment by an agent with identifier id. The virtual environment of in the vehicle routing system  does not provide virtual actions. 

An example of an external action in the vehicle routing system is an vehicle agent that advises the driver about a preferable route. An example in the AGV transportation system is an AGV agent that commands the vehicle to pick a load:

\footnotesize{
\begin{verbatim}
    public Pick(String name, Node node) { ... }
    ...
    pick = new Pick("pick", n45);  
    virtualEnv.invoke(id, pick);  
\end{verbatim}
}
\normalsize

The pick action consists of a name and the node where the load is located. Node is an abstraction of a position on the map of the environment, maintained by the agents. 
This action is then then converted to an operation for Egemin's AGV steering system E'nsor\textsuperscript{\textregistered} (Egemin Navigation System On Robot). For an AGV with a single fork lift, the pick action only requires the segment where the load is located.  The physical execution of the action, such as staying on track, turning, and the manipulation of the load is handled by E'nsor. Actually, the control commands provided by the E'nsor interface are coded in a low-level digital format. The virtual environment therefore uses the underlying E'pia\textsuperscript{\textregistered} middleware (Egemin Platform for Integrated Automation) to invoke external action, which translates the operations to E’nsor control commands.


Similarly, to inform a driver about a route in the vehicle routing system, the action is translated to interface commands for a dashboard driver. To support such external actions, the action service can provide supporting functionality to translate action requests to low-level operations.  Alternatively, supporting middleware services are available that support action translation.  
\vspace{5pt}\newline
The \textbf{Communication Service} is responsible for managing the communicative interactions among agents. The service receives messages, provides the necessary infrastructure to buffer messages, and deliver messages to the appropriate agents. Communication service is defined as follows:

\footnotesize{
\begin{verbatim}
Component CommunicationService is abstraction ()
{
     type Message is view[id : Integer, sender : String, receiver : String, 
            protocol : String, performative : String, content : Any].
     type Transmission is Any.
     ...

     port communicate is { connection sendMsg is in(Message).
                           connection receiveMsg is out(Message) }.
     port transmit is { connection transmission is out(Transmission).
                        connection deliver is in(Transmission) }.
     ...

     addresses is location(set(StateItems)).

     behavior is {
         analyzeMessage is function ( msg : Message ) :
           ItemStates { unobservable }.
         localCommunication is function( msg_items: StateItems,
           addrs : location(set(StateItems) ) :
             Boolean { unobservable }.
         generateTransmission is function( msg_items : StateItems,
           addrs : location(set(StateItems) ) :
             Transmission { unobservable }.
         analyzeTransmission is function( trans : Transmission,
           addrs : location(set(StateItems)  ) :
             StateItems { unobservable }.
         generateMessage is function( trans_items : StateItems,
           addrs : location(set(StateItems) ) :
             Message { unobservable }.
         ...
         choose
         {
            //send message
            unobservable.
            via sendMsg receive message_in.
            message_items := analyzeMessage(message_in).
            if localCommunication(message_items, addresses)
                 then //deliver message locally
                      via receiveMsg send message_in.
                      behavior().
                 else //deliver message remotely
                      transmission_out =
                        generateTransmission(message_state, addresses).
                      via transmission send transmission_out.
                      behavior().
         or
            //deliver message
            unobservable;
            via deliver receive transmission_in.
            transmission_data :=
              analyzeTransmission(transmission_in, addresses).
            message_out := generateMessage(transmission_data, addresses).
            via receiveMsg send message_out.
            behavior().
        }
}
\end{verbatim}
}
\normalsize
An agent communication message consists of message identifier, the names of the sender and receiver, the message performative (inform, request, propose, etc.), the name of the protocol, 
and the content of the message. The content is described in a content language which is based on a shared ontology. The ontology defines a vocabulary of words that enables agents to refer unambiguously to concepts and relationships between concepts in the domain when exchanging messages. Such message descriptions enable a designer to express the communicative interactions
independently of the underlying communication technology.

When the message service receives a message, it analyzes the message and checks whether it concerns a message for a local or a remote addressee.  In the former case, it forwards the message to the right agent, in the latter case a transmission is generated and the message is transmitted to the remote agent.  To actually transmit the messages, the communication service makes use of a distributed communication system provided by underlying middleware. The communication service can provide supporting functions for exchanging messages in a distributed setting, such as white and yellow page services, or alternatively, the service may depend on the underlying middleware for such functions.  

%
When the communication service receives a transmission from  a remote host, it analyses it and generates a message from the received data in the agent message format. The message is then forwarded to the agent.  

In the AGV transportation systems, we developed two approaches for task assignment, one based on computational fields (we explain this approach below) and another approach in which agents exchange messages~\cite{FBvsPB}. In the message-based approach, transport agents and AGV agents exchange messages with one another to agree on the assignment of transportation tasks. 
Here is an example of a message for a call for proposals sent by a transport agent: 

\footnotesize{
\begin{verbatim}
    public Message(int id,  String sender, String receiver, 
        String protocol, String performative, Content c) { 
        ... 
    }
    ...
    transport = new Content("regular", "5", "n60");
    cfp = new Message(77, "ta14", "20m", "DynCNET", "cfp", transport);  
    virtualEnv.send(id, cfp);  
\end{verbatim}
}
\normalsize

The identifier of this message is 77. The message is sent by transport agent ta14. Receivers are all AGV agents within a scope of 20 m from the location of the transport. The protocol is DynCNET (dynamic Contract Net~\cite{FBvsPB}) and the performative of the message is cfp (call for proposals). We elaborate on the DynCNET protocol when we discuss the protocol-based communication pattern.  The content of the message contains 3 elements:  regular, i.e., the type of the transport; 5, i.e., the priority of the transport; and n60, i.e., the location of the transport. The receivers use the performative to parse the content of the message correctly.  

An example of a message in the vehicle routing system is a vehicle agent that sends a reservation ant (i.e., a smart message) to book a particular path along the traffic network. 

\footnotesize{
\begin{verbatim}
    public Message(int id,  String sender, String receiver, 
        String protocol,  String performative, Content c) { 
        ... 
    }
    ...
    booking = new Content(...);
    ant = new Message(44, "va174", "ia75", 
           "PropagateIntention", "booking", booking);  
    virtualEnv.send(ant);  
\end{verbatim}
}
\normalsize

This message with identifier 44 is sent by vehicle agent va147. The first receiver is infrastructure agent ia75. The name of the protocol is propagate intention. The performative says that the message is a booking. The booking contains a sequence of entries that correspond with road segments along the path of the booking, including the names of the infrastructure agents along the path, and timing data. When an infrastructure agent receives the message, it parses the content to extract the data to book the local reservation. Then the agent acknowledges its reservation by marking the entry of its booking in the content of the message. Finally, the agent updates the sender with its id and the receiver with the id of the next infrastructure agent in the list, and forwards the messages. Once the infrastructure agent at the final destination has made its booking it sends the booking message back to the vehicle agent to notify the agent about the booking. 
\vspace{5pt}\newline
\textbf{Synchronization} is responsible for synchronizing state of the virtual environment with external entities, i.e., virtual environments on neighboring nodes as well as external resources.
The $\pi$-ADL specification of synchronization is defined as follows:

\footnotesize{
\begin{verbatim}
component Synchronization is abstraction ()
{
     type SynchronizationUpdate is Any.
     ...

     port synchronize is { connection syncOut is out(SynchronizationUpdate).
                           connection syncIn is in(SynchronizationUpdate) }.
     ...

     synchronizationItems is location( set(StateItems) ).

     behavior is {
         selectItems is function( item : location(set(StateItems)) :
           StateItems { unobservable }.
         localSync is function(  state : StateItems,  
           location( set(StateItems) )) : 
            Boolean { unobservable }.
         ... 
         generateSyncUpdate is function( state : StateItems ) :
           SynchronizationUpdate { unobservable }.
         decodeSyncUpdate is function( update : SynchronizationUpdate) :
           StateItems { unobservable }.

         choose
         {
            //locally initiated synchronization
            unobservable.
            via stateTemplate send selectItems(synchronizationItems).
            via readState receive state_items : StateItems.
            if localSync(state_items, synchronizationItems)
               //local state update
               via writeState send updateState(state_items).
            else
               //external synchronization
               unobservable.
               via syncOut send generateSyncUpdate(state_items).
            behavior().
         or
            //remotely initiated synchronization
            unobservable.
            via syncIn receive sync_update : SynchronizationUpdate.
            via writeState send decodeSyncUpdate(sync_update).
            behavior().
        }
     }
}
\end{verbatim}
}
\normalsize

A synchronization update defines a synchronization interaction with the external environment. This abstract type has to be instantiated for the domain at hand. A synchronization can be initiated either locally or remotely. In the former case, a distinction can be made between a synchronization that updates the local state of the virtual environment, and one that triggers an external update. In the latter case, the received synchronization update is decoded and the resulting data is used to update the local state repository. 

An example of synchronization in the AGV transportation system is the coordination required to resolve conflicts between path projections requested by AGV agents. To realize this coordination, the synchronization components of the AGV virtual environments rely on the ObjectPlaces middleware~\cite{SchePhD}. We developed ObjectPlaces to encapsulate the tedious management tasks associated with distribution in mobile applications, and applied it to the AGV transportation system. ObjectPlaces enables to set up an automatically up-to-date collection of data objects on a set of nodes in the network, and maintain these objects in the face of dynamically changing conditions. To resolve conflicts between path projections requested by different AGV agents, ObjectPlaces uses a mutual exclusion protocol. For a detailed discussion of path projection coordination with ObjectPlaces, we refer the interested reader to~\cite{Sche:Wey:HolDSO}. 

\begin{figure}[ht!]
\centering
\resizebox{\textwidth}{!}
{\includegraphics{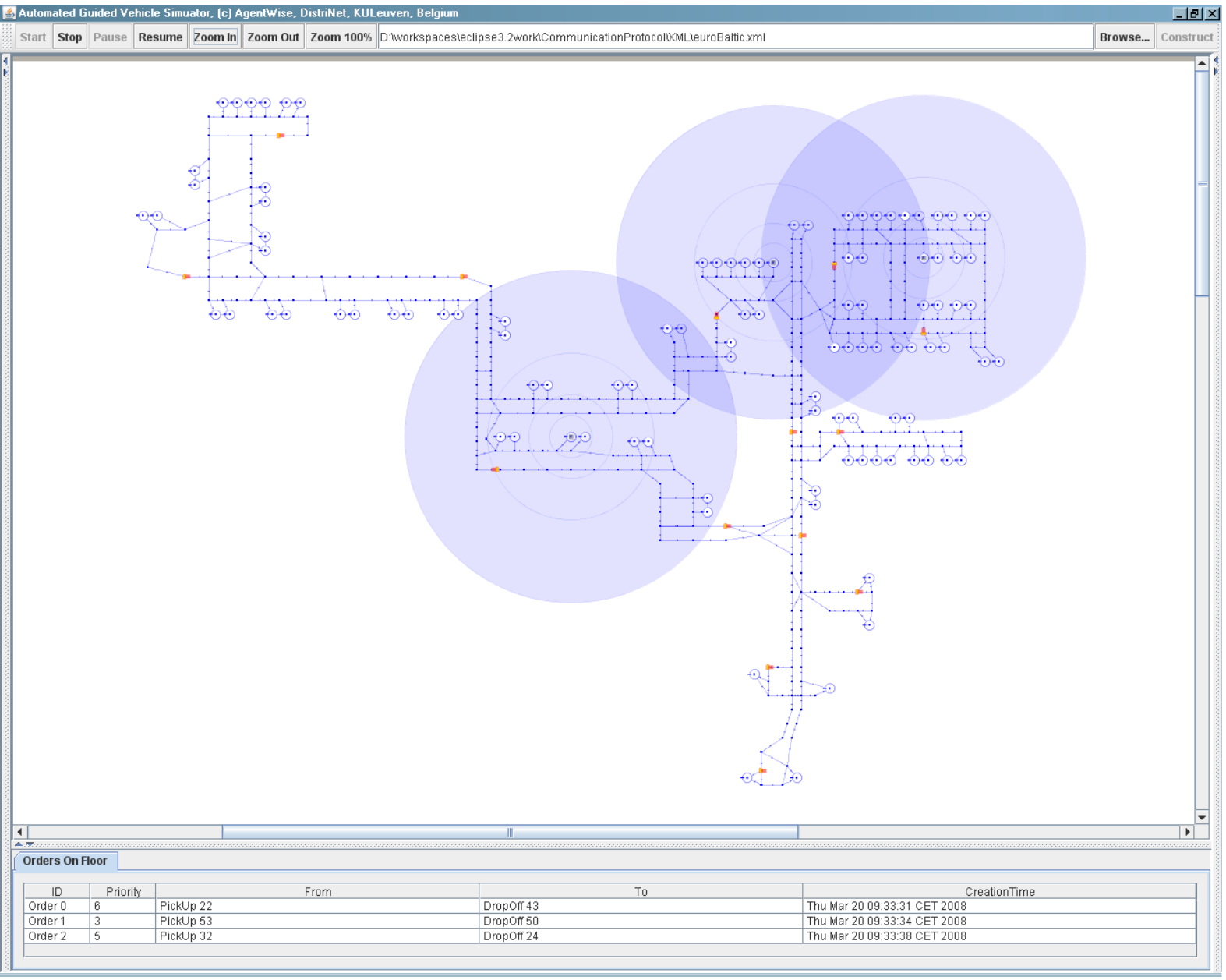}}
\caption{Transport fields emitted in the virtual environment}
\label{fig:fields}
\end{figure}

We illustrate synchronization in the AGV transportation with another example. As explained above, besides the message-based approach, we also developed an approach for transport assignment that is based computational fields. These fields are emitted by transport agents. The computational fields are distributed data structures spread in through the virtual environment, which attract AGV agents. Each AGV agent of an idle AGV combines the sensed fields and follows the gradient of the combined field, guiding its AGV to a load~\cite{FBvsPB}. Fig.~\ref{fig:fields} shows a snapshot from the simulation of an industrial AGV transportation system in which a number of transport agents emit a field in the virtual environment. The range of the fields is variable and depends on the priority of the tasks. In this case, the synchronization components of the local virtual environments are responsible for maintaining the fields.

This field-based approach employs both types of locally initiated synchronizations. In particular, if a task is not assigned after a predefined amount of time, its priority is increased. Such update is triggered locally by the synchronization component and updates the local state of the field in the virtual environment. On the other hand, a change in priority of a field triggers the synchronization component to spread of field in the virtual environment to neighboring nodes. This is an example of a locally initiated synchronization that triggers an external update. 

\footnotesize{
\begin{verbatim}
    public class FieldData {
        private int id; 
        private int priority;
        private Node source;  
        ...
    }

    public class TaskField {
        private Task task;
        private FieldData fieldData;
        ...
    }
    
    ...
    //transport agent creates a field and then performs an emit action 
    fdata = new FieldData(120, 3, n60); 
    field = new TaskField(t60, fdata); 
    emit = new Emit(`emit', field);  
    virtualEnv.invoke(ta8, emit);  
    
    //virtual environment distributes the field
    objectplace.put(field, callback);  
    ...
\end{verbatim}
}
\normalsize

The excerpt shows how transport agent ta8 located at node n60 creates a field with id 120 and priority 3 for transportation task t60, and then emits the field in the virtual environment. Subsequently, the virtual environment distributes the field to other nodes using  the \textit{put} operation provided by the underlying ObjectPlaces middleware (the put operation is an example of a domain-specific realization of a synchronization update. For emitting a field, only the nodes within a certain distance from the transport agent that emits the field will receive the field. This distance is determined based on the priority of the field (in the field data). When an AGV enters or leaves the range determined by this distance, the virtual environment of the AGV will be notified by the middleware and the field will be added or removed respectively.  When the execution of the task is not started after a certain time, the priority of the task will be increased and the virtual environment will update the field in the objectplace. This is turn will trigger the update to the other nodes via the middleware.  

An example of an externally initiated synchronization in the traffic routing application is the periodic update of the position of a car on the virtual road network. To that end, the synchronization component will use the digital data derived from the position sensor of the car to update the position of the car in the state repository of the virtual environment. 

\subsubsection{Design Rationale}\label{design-rationale-app-env}
\mbox{\ }\vspace{-6pt}\newline

Three primary principles that underlay the design of the virtual environment are separation of concerns, synchronization processes, the shared data style. 

Action service, perception service, and communication service provide operations corresponding to the various ways situated agents can access the virtual environment. By separating the various  concerns, the decomposition of the virtual environment yields a flexible modularization that can be tailored to a broad family of application domains. For example, for applications in which agents interact via marks in the virtual environment but do not communicate via message exchange, the communication service can be omitted. 

The synchronization component contributes to the flexibility and openness of the multi-agent system. Synchronization can encapsulate processes that maintain a representation of the state of external resources in the state repository, which reflect dynamics in the environment. Agents can perceive these dynamics, which allow them to flexibly adapt their behavior when operating conditions change. Furthermore, synchronization processes can keep track of agents that come and go and reflect this in the state repository of the virtual environment. This indirection reduces the coupling between agents enhancing the openness of the multi-agent system. 

The shared data style results in low coupling between elements that have clear and disjunctive responsibilities. This makes the elements better understandable in isolation. Furthermore, it improves reuse (elements are not dependent on too many other elements) and modifiability (changes in one element do not affect other elements or the changes have only a local effect).

\subsection{Situated Agent}

\subsubsection{Primary Presentation}
\mbox{\ }\vspace{-6pt}\newline

\begin{figure}[h!]
\centering
\resizebox{.8\textwidth}{!}
{\includegraphics{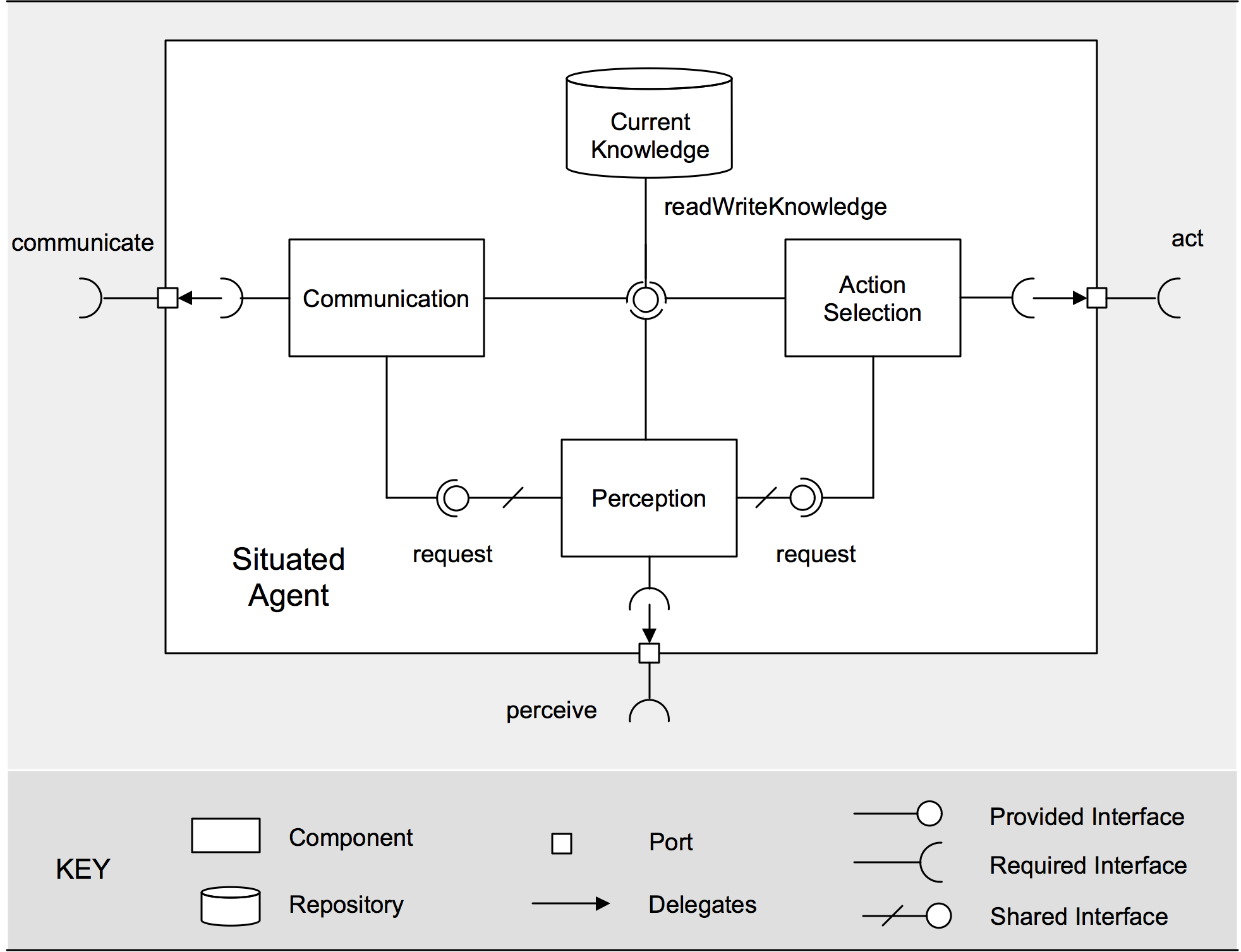}}
\caption{Primary presentation of situated agent}
\label{fig:situatedagent}
\end{figure}

The primary presentation of the situated agent pattern is shown in Fig.~\ref{fig:situatedagent}. 

A situated agent is an autonomous entity that is situated in a virtual environment. A situated agent encapsulates its state and controls its behavior, aiming to realize the tasks it is assigned. Situated agents are able to adapt their behavior according to changing operating conditions. A situated agent is a cooperative entity. The realization of the overall tasks results from interaction among agents. 
At the top level, the $\pi$-ADL specification of situated agent is defined as follows:

\footnotesize{
\begin{verbatim}
component SituatedAgent is abstraction ()
{
   //type definitions
   type Focus is view[focusName : String, focusParams : Any].
   type Sense is view[agentId : String, focus : Focus].
   type Representation is Any.
   type Action is view[agentId : String, actionName : String,
            actionParams : Any].
   type Message is view[id : Integer, sender : String, receiver :
            protocol : String, performative : String, content : Any].

   port perceive is { connection sense is out(Sense).
                      connection sensed is in(Representation) }.
   port act is { connection action is out(Action) }.
   port communicate is { connection send is out(Message).
                         connection receive is in(Message) }.

   //composition
   behavior is compose {
            perception is Perception()
       and  communication is Communication()
       and  actionSelection is ActionSelection()
       and  currentKnowledge is CurrentKnowledge()
   }  where {
            perception::sense relays sense
       and  perception::request unifies communication::request
       and  perception::request unifies actionSelection::request
       and  sensed relays perception::sensed
       and  communication::send relays send
       and  receive relays communication::receive
       and  actionSelection::action relays action
       and  currentKnowlege::readWriteKnowledge unifies 
               communication::readWriteKnowledge
       and  currentKnowlege::readWriteKnowledge unifies 
               actionSelection::readWriteKnowledge
       and  currentKnowlege::readWriteKnowledge unifies 
               perception::readWriteKnowledge
   }                			
}
\end{verbatim}
}
\normalsize

The specification starts with defining types and ports of a situated agent. Then, the composition of the constituent components of a situated agent is defined. We introduce these components in the following subsection. An elaborated discussion of perception, action selection and communication follows in the next sections that discuss refined patterns for each of these components. 

\subsubsection{Architectural Elements}
\mbox{\ }\vspace{5pt}\newline
%
The \textbf{Current Knowledge} repository contains data that is shared among the components of a situated agent. The $\pi$-ADL specification of current knowledge is defined as follows:

\footnotesize{
\begin{verbatim}
component CurrentKnowledge is abstraction()
{
   type KnowledgeItem is view[name : String, val : Any].
   type Knowledge is set(KnowledgeItem).

   port readWriteKnowledge is { 
                       connection knowledgeTemplate is in(Knowledge).
                       connection readKnowlege is out(Knowledge). 
                       connection writeKnowledge is in(Knowledge) }.
          assuming { protocol is (via knowledgeTemplate receive Knowledge.
                                  true*.via readState send Knowledge.)* 
                                  or (via writeState receive Knowledge.)* }.

   currentKnowledge is location(set(Knowledge)).

   behavior is {
      write is function( knowledge : Knowledge,
        repository : set(Knowledge) ) :
          location(set(Knowledge)) { unobservable. }.
      read is function( template : Knowledge,
        repository : set(Knowledge) ) :
          Knowledge { unobservable }.

      choose
      {
         //read knowledge
         	via knowledgeTemplate receive knowledge_template : Knowledge.
          unobservable.
         	via readKnowledge send read(knowledge_template, currentKnowledge).
         	behavior().
      or
         //write knowledge
         	via writeKnowledge receive knowledge_items : Knowledge.
          unobservable.
         	currentKnowledge := write(knowledge_items, currentKnowledge).
         	behavior().
      }
   }
}
\end{verbatim}
}
\normalsize

Knowledge items are abstract elements that represent something of interest to the agent. Such items can refer to elements of the external environment, elements in the virtual environment, or internal elements. An agent's current knowledge is maintained in a repository that provides a shared port that allows components to read and write knowledge. The protocol of the read write port enforces the atomic completion of either reading or writing the repository. Reading knowledge requires a template and after processing the requested knowledge is returned. Reading and writing are further specified in the behavioral part of the definition. 

Examples of knowledge in the vehicle routing system are a path along the road network represented as a graph of nodes (see the definition of graph path in section~\ref{sec:arch-elems-ve} where we discuss the perception service), the current position of the car as a node of the graph, the status of the traffic in the monitored area, bookings for road segments, etc. 
An example of knowledge in the AGV transportation systems is operating space that an AGV agent uses to keep track of its path projections in the virtual environment. 

\footnotesize{
\begin{verbatim}
  public class OperatingSpace {

     private Path requestedPath;
     private Path lockedPath;
     ...

     public void locked(Path p)  { ... } 
     public void release(Path p)  { ... } 
     ...
  }
\end{verbatim}
}
\normalsize

An operating space keeps track of a requested path and a locked path. The requested path has been projected in the virtual environment, but is not locked yet (at least the AGV agent has not observed this). The locked path is reserved for the AGV to move safely. Operation space provides a lock operation to add a projected path to the locked path once it is locked, and an operation to release the  path behind the AGV when it moves. 
\vspace{2pt}\\
\textbf{Perception} is responsible for collecting data for the agent. This data concerns either state of the virtual environment or state of the environment. The $\pi$-ADL specification of perception is defined as follows:

\footnotesize{
\begin{verbatim}
component Perception is abstraction()
{
   type Focus is view[focusName : String, focusParams : Any].
   type Filter is view[filterName : String, filterParams : Any].
   type PerceptionRequest is view[id : String, focus : Focus, filter : Filter].
   type Sense is view[agentId : String, focus : Focus].
   type Representation is Any.
   ...

   port request is { connection requestPerception is in(PerceptionRequest).
                     connection notifyRequest is out(String)}.
   port perceive is { connection sense is out(Sense).
                      connection sensed is in(Representation) }.
   ...

   behavior is {
      generateSense is function( p_request : PerceptionRequest) :
          Sense { unobservable }.

      via requestPerception receive request : PerceptionRequest.
      via sense send generateSense(request). 
      unobservable. 
      via sensed receive representation : Representation. 
      unobservable.
      via writeKnowledge send update : Knowledge.
      via notifyRequest send request::id. 
      behavior().
   }
}
\end{verbatim}
}
\normalsize

The perception component enables an agent to sense the virtual environment for information.
To support selective perception, a perception request includes a selected focus and filter. We explained the concept of a focus in section~\ref{sec:arch-elems-ve}. A filter allows an agent to select only those elements of a representation that match a particular selection criteria. 

 Perception requests can be triggered by the communication component and the action selection component. When perception receives a request, it generates a sense that is used to sense the virtual environment, resulting in a representation. Perception processes the representation, updates the agent's current knowledge, and then notifies the perception requester.  

%
An example of a filter in the AGV transportation system is a filter to select the nearest transportation task that resulted from sensing the tasks within a certain distance. An example in the vehicle routing system is a filter to select the shortest path from a set of sensed paths:

\footnotesize{
\begin{verbatim}
    public ShortestPath(Sting name, GraphPaths paths) {
       ...
    }
    pfilter = new ShortestPath(`shortestPath', sensedPaths); 
\end{verbatim}
}
\normalsize

This filter selects the shortest path from a given set of sensed paths. The representation of graph paths is discussed in section~\ref{sec:arch-elems-ve}. We elaborate on the perception when we discuss the selective perception pattern.  
%
\vspace{4pt}\newline
\textbf{Behavior-Based Action Selection}
is responsible for realizing the agent's objectives by invoking actions in the virtual environment. Actions either intend to manipulate the state of the virtual environment or manipulate external resources. The $\pi$-ADL specification of action selection is defined as follows:

\footnotesize{
\begin{verbatim}
component ActionSelection is abstraction()
{
    type Action is view[agentId : String, actionName : String,
             actionParams : Any].
    ...

    port act is { connection action is out(Action) }.

    behavior is {
      unobservable.  
      via action send action : Action.     
      behavior().
   }
}
\end{verbatim}
}
\normalsize

After the definition of types and ports, the abstract behavior of action selection is defined. During internal processing, action selection can decide to select an action and subsequently, the selected action is invoked  in the virtual environment. 

The following excerpt illustrates how an action is realized in the vehicle routing system:  

\footnotesize{
\begin{verbatim}
    public InstructDriver(String name, GraphPath path) { 
       ... 
    }
    ...
    intention = new InstructDriver(`intention', path);  
    virtualEnv.invoke(id, intention);  
\end{verbatim}
}
\normalsize

The action advises the driver with a preferable path in the environment. For examples of actions in the AGV transportation system, we refer to the discussion of the action service in section~\ref{sec:arch-elems-ve}, where we discuss several examples.
%
%
%
We further elaborate on decision making when we discuss the behavior-based action selection pattern.
\mbox{ }\vspace{4pt}\newline
\textbf{Communication} is responsible for communicative interactions with other agents. Message exchange enables agents to share information and set up collaborations. 
The $\pi$-ADL specification of communication is defined as follows:

\footnotesize{
\begin{verbatim}
component Communication is abstraction ()
{
     type Message is view[id : Integer, sender : String, receiver : String,
            protocol : String, performative : String, content : Any].
     ...

     port communicate is { connection sendMsg is out(Message).
                           connection receiveMsg is in(Message) }.

     behavior is {
        choose
        {
          //receive message 
          via receiveMsg receive message : Message.
          unobservable.
          behavior().
        or
          //send message 
          unobservable.
          via sendMsg send message : Message.
          unobservable.   
          behavior().
        or
          //silence 
          unobservable.   
          behavior().
        }
     }
}
\end{verbatim}
}
\normalsize
After type and port declarations, the behavior of communication is defined. The communication component either receives a message that is internally processed, or it sends a message after internal processing, or its behavior is silent (e.g., a conversation is terminated without further notice). We have presented concrete realizations of messages for both the vehicle routing system and the AGV transportation system in section~\ref{sec:arch-elems-ve}, where we discuss the communication service. 
%
%
We elaborate on communication and give a concrete example of a protocol when we explain the protocol-based communication pattern.

\subsubsection{Design Rationale}\label{design-rationale-app-env}
\mbox{\ }\vspace{5pt}\newline
The primary principles that underlay the design of situated agent are decentralized control and locality, perception on demand, and parallelism of action selection and communication. 




In a situated multi-agent system, control is decentralized and divided among the agents. Situated agents manage the dynamic and changing operating conditions locally and autonomously, enabling them to flexibly adapt their behavior when needed. 
However, decentralized control (the system tasks are realized by agents that make local decisions) typically requires more communication. The performance of the system may be affected by the communication links between agents, which could impose a major bottleneck. Furthermore, agents' decision making is based on local information, which may result in suboptimal system behavior. These tradeoffs and should be kept in mind throughout the design and development of a situated multi-agent system. 

Selective perception enables an agent to focus its attention to the relevant aspects in the environment according to its current tasks. When selecting actions and communicating messages with other agents, decision making and communication typically request perceptions to update the agent’s knowledge. By selecting an appropriate focus and filter, the agent directs its attention to the current aspects of its interest, and adapts it attention when the operating conditions change.

The overall behavior of a situated agent is the result of the coordination of action selection and communication. Action selection is responsible for selecting suitable actions to perform tasks. Communication is responsible for the communicative interactions with other agents. However, the two components coordinate to complete the agent's tasks more efficiently. For example, agents can send each other messages with requests for information that enable them to act more purposefully, or support each other in a collaboration. Ensuring that both action selection and communication behave in a coordinated way requires a careful design. On the other hand, the separation of functionality for coordination (via communication) from the functionality to perform actions to complete tasks has several advantages. Besides improving reusability, it allows both functions to act in parallel and at a different pace. In many applications, sending messages and executing actions happen at different tempo. A typical example domain is robotics, but it applies to any application in which the time required for performing actions in the environment differs significantly from the time to communicate messages. Separation of communication from performing actions enables agents to reconsider the coordination of their behavior while they perform actions, improving adaptability and efficiency.


\subsection{Selective Perception}

\subsubsection{Primary Presentation}
\mbox{\ }\vspace{-6pt}\newline

\begin{figure}[h!]
\centering
\resizebox{0.8\textwidth}{!}
{\includegraphics{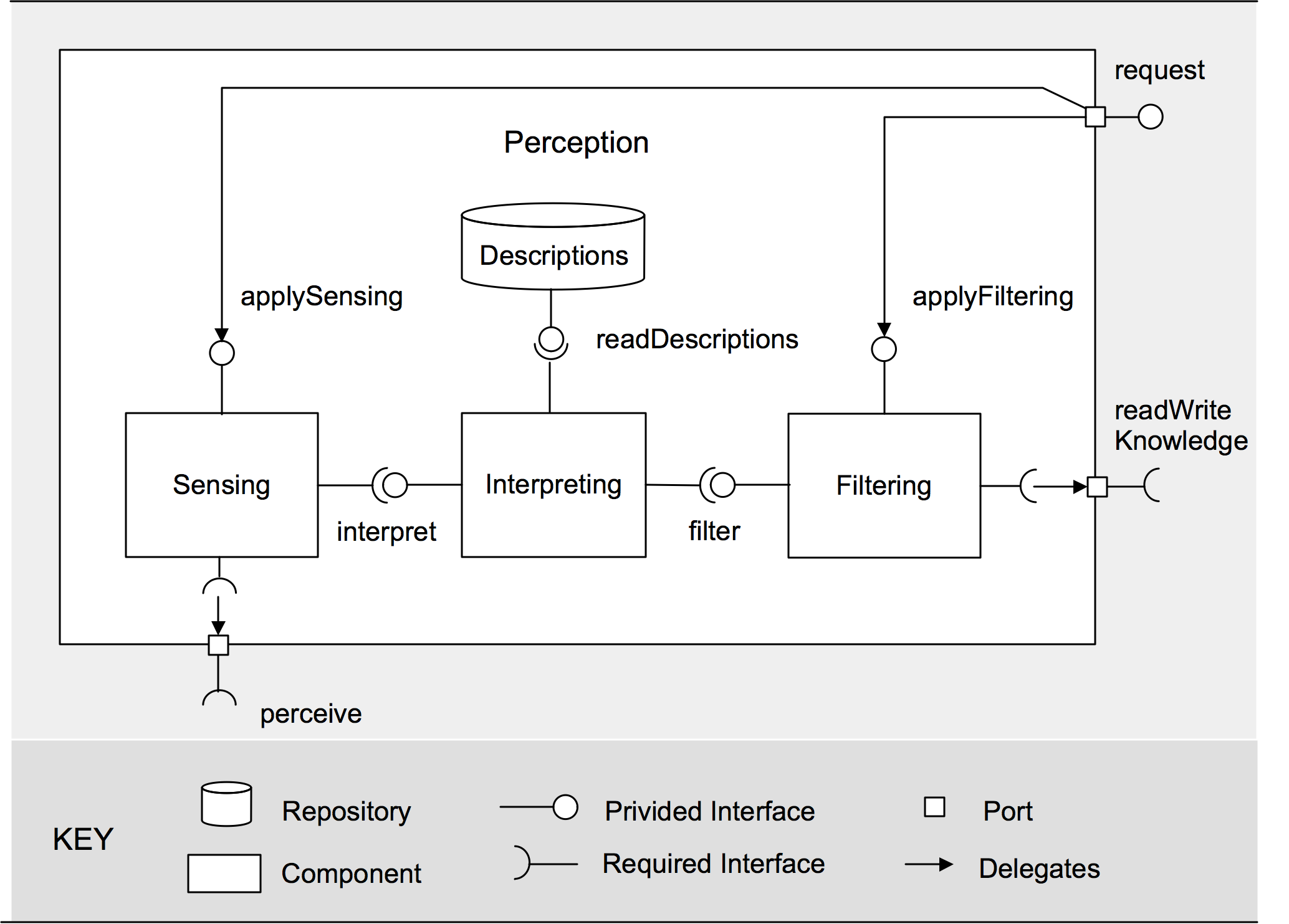}}
\caption{Primary presentation of selective perception}
\label{fig:selective-perception}
\end{figure}

The primary presentation of the selective perception pattern is shown in Fig.~\ref{fig:selective-perception}.

Selective perception is responsible for collecting runtime information to update the agent's current knowledge. Perception requests can be initiated by the communication component and the action selection component. The $\pi$-ADL specification of communication is defined as follows:

\footnotesize{
\begin{verbatim}
component Perception is abstraction()
{
   type Focus is view[focusName : String, focusParams : Any].
   type Filter is view[filterName : String, filterParams : Any].
   ...

   port perceive is { connection sense is out(Sense).
                      connection sensed is in(Representation) }.
   port request is { connection requestPerception is in(PerceptionRequest) 
                      connection notifyRequest is out(String) }.
   port readWriteKnowledge is {
          connection knowledgeTemplate is out(Knowledge).
          connection readKnowledge is in(Knowledge).
          connection writeKnowledge is out(Knowledge) }.

   behavior is compose {             
            sensing is Sensing().
        and interpreting is Interpreting().
        and descriptions is Descriptions().
        and filtering is Filtering().
   } where {
            requestPerception relays sensing::applySensing
        and sensing::sense relays sense
        and sensed relays sensing::sensed
        and sensing::interpret unifies interpreting::interpret
        and interpreting::readDescriptions unifies 
              descriptions::readDescriptions
        and interpreting::filter unifies filtering::filter
        and requestPerception relays filtering::requestPerception
        and filtering::readWriteKnowledge relays readWriteKnowledge
        and filtering::notifyRequest relays notifyRequest
      }                			
}
\end{verbatim}
}
\normalsize

After type and port declarations, the composition of the constituent components of perception is defined. We now discuss these components further in detail.
 
\subsubsection{Architectural Elements}
\mbox{\ }\vspace{-6pt}\newline

The perception components comprises one data repository: descriptions, and three components: sensing, interpreting, and filtering. To explain the various elements, we follow the logical thread of successive activities that take place from the moment a requester takes the initiative to sense the virtual environment until the agent's current knowledge is updated and the requester is notified.
\vspace{4pt}\\
\textbf{Sensing} is defined as follows: 

\footnotesize{
\begin{verbatim}
component Sensing is abstraction()
{
   ...
   type PerceptionRequest is view[id : String, focus : Focus, filter : Filter].
   type Sense is view[agentId : String, focus : Focus].
   type Representation is Any.
   ...
   port perceive is { connection sense is out(Sense).
                      connection sensed is in(Representation) }.
   port apply is { connection applySensing is in(PerceptionRequest) }.
   port interpret is { connection percept is out(Representation) }.

   behavior is {
      generateSense is function(request : PerceptionRequest) : 
          Sense  { unobservable. }.

      via applySensing receive request : PerceptionRequest.
      via sense send generateSense(request).
      unobservable.
      via sensed receive representation : Representation.
      unobservable.
      via percept send representation.
      behavior().
   }
}
\end{verbatim}
}
\normalsize

Sensing is initiated by a perception request with a focus. Sensing applies sense requests in the virtual environment and sends the resulting representation to the interpreting component. We have explained and illustrated the concepts of focus and representation in the discussion of the virtual environment pattern, see section~\ref{sec:arch-elems-ve}.
\vspace{4pt}\\
\textbf{Descriptions} is defined as follows: 

\footnotesize{
\begin{verbatim}
component Descriptions is abstraction()
{
   type Representation is Any.
   type Description is Any.

   port readDescriptions is { 
          connection representation is in(Representation).
          connection description is out(set(Description)) }
        assuming { protocol is { (via representation receive Any.true*.
                   via description send Any.)* } }.   

   descriptions is location(set(Description)).

   behavior is {
      select is function(rep : Representation, des : set(Description) ) :
        set(Description) { unobservable. }.

      unobservable.	 	
      via representation receive repres : Representation.
      via description send select(repres, descriptions).
      behavior().
   }
}
\end{verbatim}
}
\normalsize

%

Description maintains a repository of \emph{descriptions} to interpret representations.
A description specifies a particular pattern of the data contained in a representation. An example in the vehicle routing system  is a description used by infrastructure agents to map a representation of the traffic state to a level of traffic congestion in a monitored range:

\footnotesize{
\begin{verbatim}
    public  class TrafficObservation { 
       private double density;
       private double intensity;
       private double averageSpeed;
       ... 
    }    
    public class CongestionLevel { 
       private TrafficObservation observation: 
       ... 
       public CongestionLevel(TrafficObservation o)  { 
          ...
       }
    }
\end{verbatim}
}
\normalsize

Traffic observation is a representation of the current traffic in a monitored range. This representation comprises three traffic variables in the monitored range: the current density (i.e., the number of vehicles per length unit), intensity (i.e., the number of vehicles per time unit) and average speed. Given a traffic representation, the congestion level description allows to determine the current the traffic state in the monitored range.
\vspace{4pt}\\
\textbf{Interpreting} is defined as follows:

\footnotesize{
\begin{verbatim}
component Interpreting is abstraction()
{
   type Representation is Any.
   type Description is Any.
   type Knowledge is set(KnowledgeItems).
   ...

   port interpret is { connection interpretIn is in(Representation) }.
   port readDescriptions is { 
             connection representation is out(Representation).
             connection description is in(set(Description)) }.
   port filter is { connection interpretOut is out(Knowledge) }.

   behavior is {
      interpreting is function(r : Representation, ds : set(Description)) :
        Knowledge { unobservable. }.	 	
      via interpretIn receive repres : Representation.
      via representation send repres.
      unobservable. 
      via description receive descriptions : Description.
      via interpretOut send interpreting(repres, descriptions).
      behavior().
   }
}
\end{verbatim}
}
\normalsize

The interpreting component uses a set of descriptions to extract knowledge from a representation. 
This knowledge represent elements sensed in the virtual or the external environment that can be used to update the current knowledge of the agent. 

An example in the traffic monitoring system is the traffic state derived from a traffic observation using a congestion level description: 

\footnotesize{
\begin{verbatim}
    public class TrafficState { 
       private GraphPath path;  
       private Status status;
       ... 
    }
\end{verbatim}
}
\normalsize

The path represents the part of the road network that is monitored by an infrastructure agent. Status is the actual status of the traffic on the path, ranging from free flow to jammed. The infrastructure agent employs the traffic state to provide predictive traffic intensity information which is used to improve bookings requested by vehicle agents. For details on this prediction mechanism, we refer the interested reader to~\cite{1538945}.
%
\vspace{4pt}\\
\textbf{Filtering} is defined as follows: 

\footnotesize{
\begin{verbatim}
component Filtering is abstraction()
{
   ...
   type Filter is view[filterName : String, filterParams : Any].
   type PerceptionRequest is view[id : String, focus : Focus, filter : Filter].

   port filter is { connection filterIn is in(Knowledge) }.
   port apply is { connection applyFiltering is in(PerceptionRequest). 
                   connection notifyRequest is out(String) }.
   port readWriteKnowledge is {
          connection knowledgeTemplate is out(Knowledge).
          connection readKnowledge is in(Knowledge).
          connection writeKnowledge is out(Knowledge) }.
                     
   behavior is {
      filtering is function(percept : Knowledge, filter : Filter) :
                            Knowledge { unobservable. }.	

      unobservable.                        	
      via applyFiltering receive request : PerceptionRequest.
      unobservable.  
      via filterIn receive percept : Knowledge.
      via writeKnowledge send filtering(percept, request::filter).
      via notifyRequest send request::id. 
      behavior().
   }
}
\end{verbatim}
}
\normalsize


Filtering filters a percept using a selected filter. A filter imposes conditions on a percept that determine whether the data elements of the percept can pass the filter or not. The filtering component uses the filtered percept to update the agent's current knowledge, and then notifies the requester. 
We discussed an example of a filter in the vehicle routing system that selects the shortest path from a given set of sensed paths. Here is an example in the AGV transportation system: 

\footnotesize{
\begin{verbatim}
    public ParkLocation(Sting name, Node pos, List<Node> locations) {
       ...
    }
    tfilter = new ParkLocation(`nearestParking', myPos, sensedLocs); 
\end{verbatim}
}
\normalsize

This filter selects the nearest park location for the given AGV position from a given set of observed park locations.   

\subsection{Behavior-Based Action Selection}

\subsubsection{Primary Presentation}
\mbox{\ }\vspace{-6pt}\newline

\begin{figure}[h!]
\centering
\resizebox{0.8\textwidth}{!}
{\includegraphics{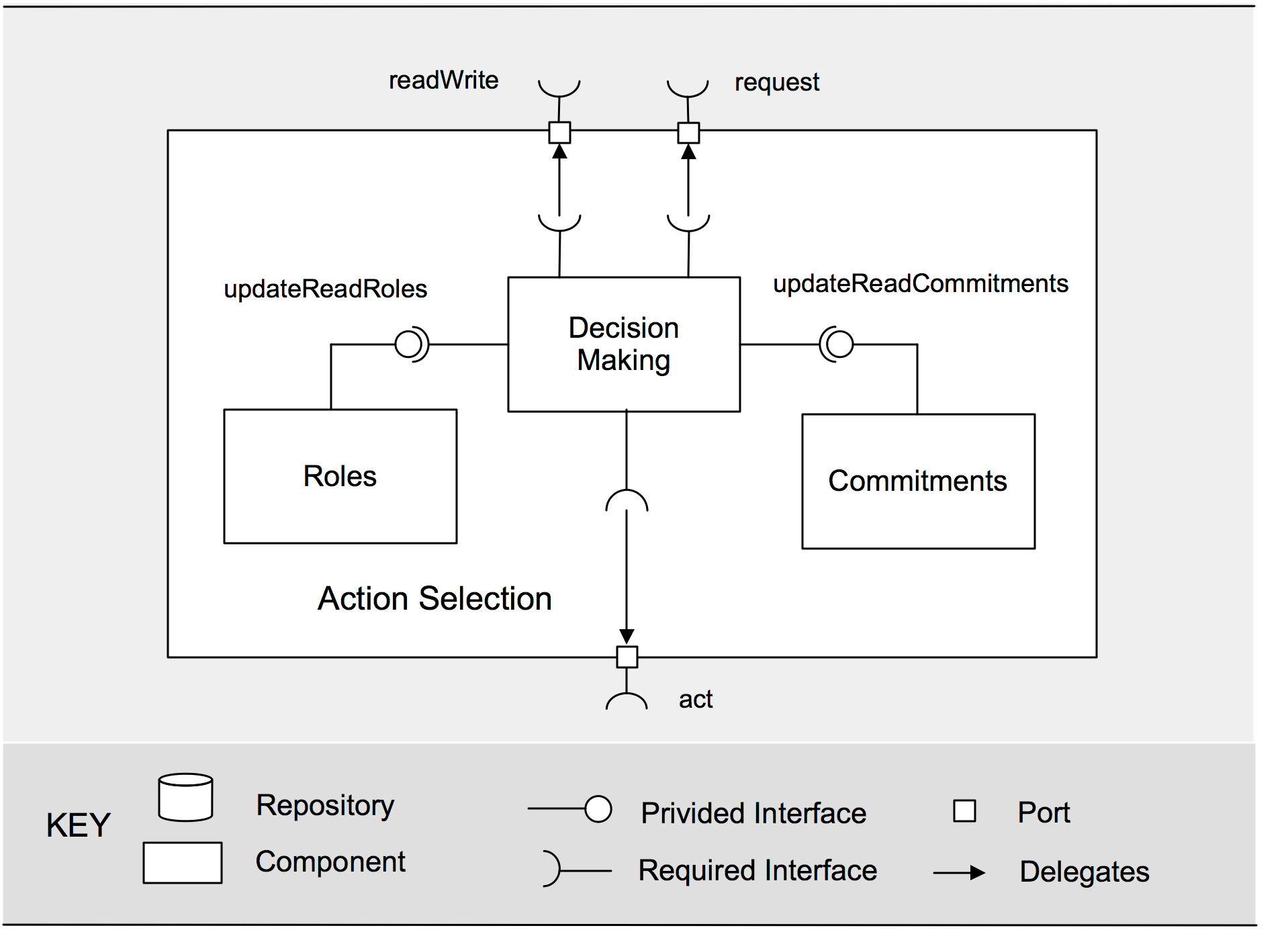}}
\caption{Primary presentation of behavior-based action selection}
\label{fig:action-selection}
\end{figure}

The primary presentation of the behavior-based action selection pattern is shown in Fig.~\ref{fig:action-selection}.

To select actions, a situated agent employs a behavior-based action selection mechanism. The main advantages of behavior-based action selection are efficiency and flexibility to deal with dynamism in the environment. The $\pi$-ADL specification of action selection is defined as follows:

\footnotesize{
\begin{verbatim}
component ActionSelection is abstraction()
{
   ...
   type Action is view[agentId : String, actionName : String, 
           actionParams : Any].

   port act is { connection invoke is out(Action) }.
   port request is { 
           connection requestPerception is out(PerceptionRequest).
           connection notifyRequest is in(String) }.
   port readWriteKnowledge is { ... }.

   behavior is compose 
   {
            decisionMaking is DecisionMaking().
        and roles is Roles().
        and commitments is SituatedCommitments().
    } where {
              decisionMaking::updateReadRoles unifies roles::updateReadRoles
        and   decisionMaking::readWriteCommitments unifies 
                 commitments::readWriteCommitments 
        and   decisionMaking::act relays act
        and   decisionMaking::knowledgeTemplate relays knowledgeTemplate
        ...
      }
}
\end{verbatim}
}
\normalsize

Central to behavior-based action selection is the decision making component that employs roles and situated commitments to select actions. A role represents a coherent part of functionality of a situated agent. Roles are relatively simple behavioral modules that tightly couple sensing to action. A situated commitment defines an engagement of an agent in a collaboration that takes into account the context of the collaboration. Decision making selects the next action of the agent based on the agent's roles and the current commitments. We discuss roles and situated commitments now in detail. 
 
\subsubsection{Architectural Elements}
\mbox{\ }\vspace{-6pt}\newline

The action selection component comprises three components: roles, situated commitments, and behavior. 
\vspace{4pt}\\
\textbf{Roles} is defined as follows: 

\footnotesize{
\begin{verbatim}
component Roles is abstraction()
{
   type Role is view[name : String, behaviors : Any, 
             actions : set(Action)].

   port updateReadRoles is { 
         connection updateRoles is in(Knowledge). 
         connection readRoles is out(set(Role)). 
   }    

   roles is set(Role).

   behavior is {
     update is function(k : Knowledge, roles : set(Role)) : 
         set(Role) { unobservable }. 
  
     choose 
    {
       //update roles      
      via updateRoles receive knowledge : Knowledge.
      unobservable.
      roles := update(knowledge, roles).     
      behavior().
    or
      //read roles 
      unobservable. 
      via readRoles send(roles).
      behavior().
   }
}
\end{verbatim}
}
\normalsize

The roles component comprises a set of roles. A role has a name, a set of behaviors, and a set of actions. The roles can be updated with new knowledge, and the set of roles can be read. 

\begin{figure}[h!]
\centering
\resizebox{.6\textwidth}{!}
{\includegraphics{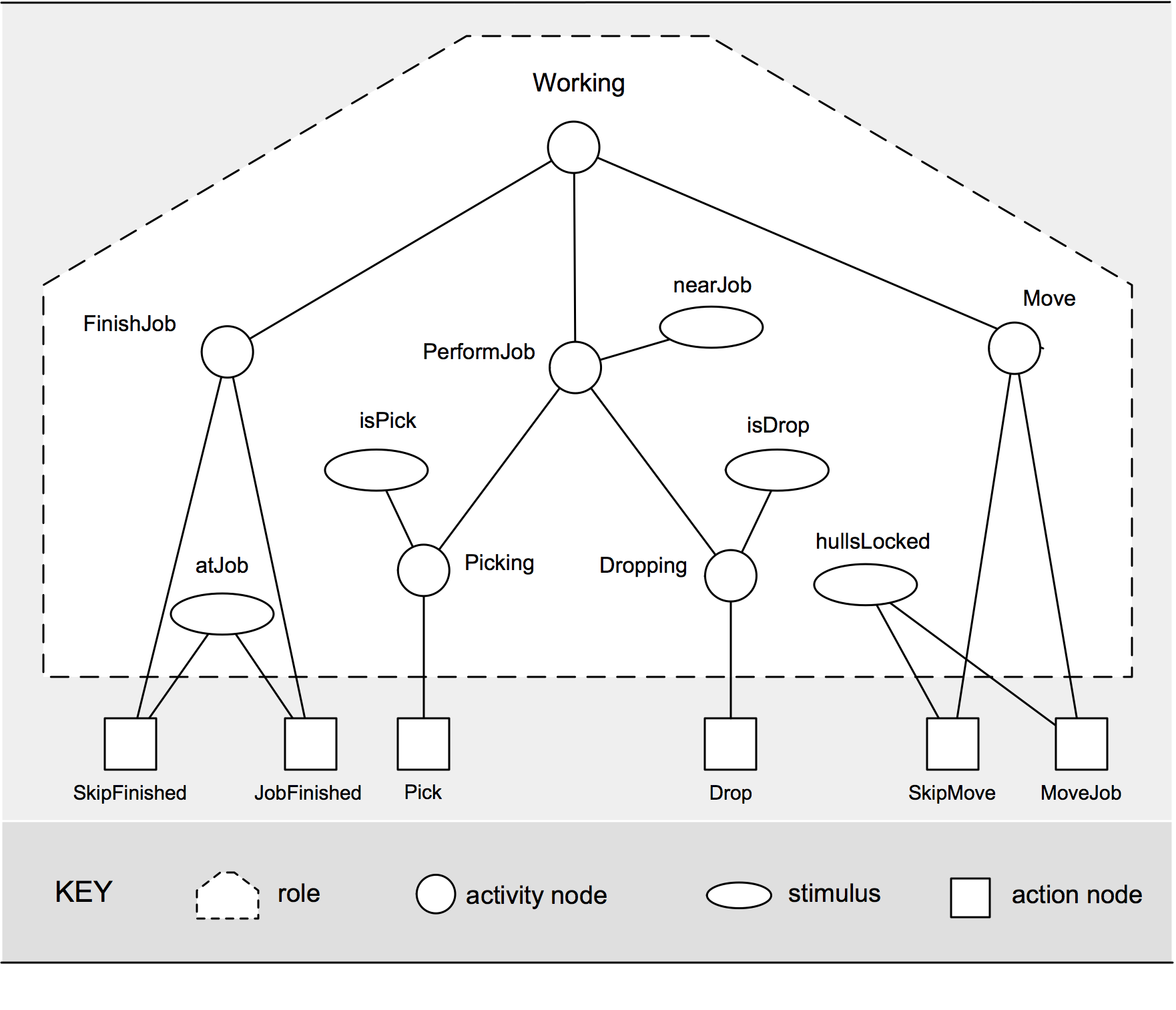}}
\caption{Design of the free-flow tree for the working role of AGV agents}
\label{fig:tree-role}
\end{figure}

The main roles of the AGV agent in the AGV transportation system are ``Working,'' ``Charging'' and ``Parking,'' which respectively denote the roles to perform a transportation task, recharge the AGV's battery, and park the vehicle when it is idle. For the concrete realization of the roles we used an extension of free-flow trees~\cite{Tyr,Bry,Stee}. Fig.~\ref{fig:tree-role} shows the design of the tree for the working role of an AGV agent. 

A free-flow tree is composed of a hierarchy of nodes representing the role behavior with leave nodes representing actions. To select an action, activity is injected at the top node of the tree. While the activity flows along the nodes additional activity may be injected based on stimuli sensed by the agent (ovals in Fig.~\ref{fig:tree-role}). When the activity arrives at the leave nodes the action with the highest activity is the preferable action for the role in the given context. 

The main behaviors of the working role are moving, performing a job (pick or drop a load), and finishing job (i.e., the actual manipulation of the load during a pick or drop). The behaviors are affected by stiumuli related to the actual position of the AGV to the pick or drop location, and the reservation of hulls to avoid collisions. Skip move is selected when no new move action is required (e.g, when the AGV waits on a crossroad), and skip finished is selected during the manipulation of a load.  
\vspace{4pt}\\
\textbf{Commitments} is defined as follows: 

\footnotesize{
\begin{verbatim}
component Commitments is abstraction()
{
   type SituatedCommitment is view[name : String, context : Any, 
             condition : Boolean, effect : Any].

   port updateReadCommitments is { 
         connection updateCommitments is in(Knowledge). 
         connection readCommitments is out(set(Commitment)). 
   }    

   commitments is set(SituatedCommitment).

   behavior is {
     update is function(k : Knowledge, cs : set(SituatedCommitment)) : 
         set(SituatedCommitment) { unobservable }. 
  
     choose 
    {
       //update commitments      
      via updateCommitments receive knowledge : Knowledge.
      unobservable.
      roles := update(knowledge, commitments).     
      behavior().
    or
      //read commitments 
      unobservable. 
      via readCommitments send(commitments).
      behavior().
   }
}
\end{verbatim}
}
\normalsize

The commitments component comprises a set of situated commitments. A situated commitment is defined by a name, a context, an activation condition, and an effect. Commitments can be updated with new knowledge, and the set of commitments can be read. Agents agree on mutual commitments in a collaboration via direct communication (see the Communication pattern below). Once the agents have agreed on a collaboration, the mutual situated commitments will affect the selection of actions in favor of the agents’ roles in the collaboration. 

An example of a situated commitment in the vehicle routing system is ``current intention'' that represents the booking of a path a vehicle agent at the infrastructure agents. This commitment will guide the driver along the selected path to the destination. The main commitments in the AGV transportation system are ``Work'' and ``Charge.'' Work is a mutual commitment between an AGV agent and a transport agent to perform a transportation task. Charge is commitment of an AGV agent to charge its battery enabling it to perform future transportation tasks. Such commitment contributes implicitly to the realization of the system tasks. 

\begin{figure}[h!]
\centering
\resizebox{.8\textwidth}{!}
{\includegraphics{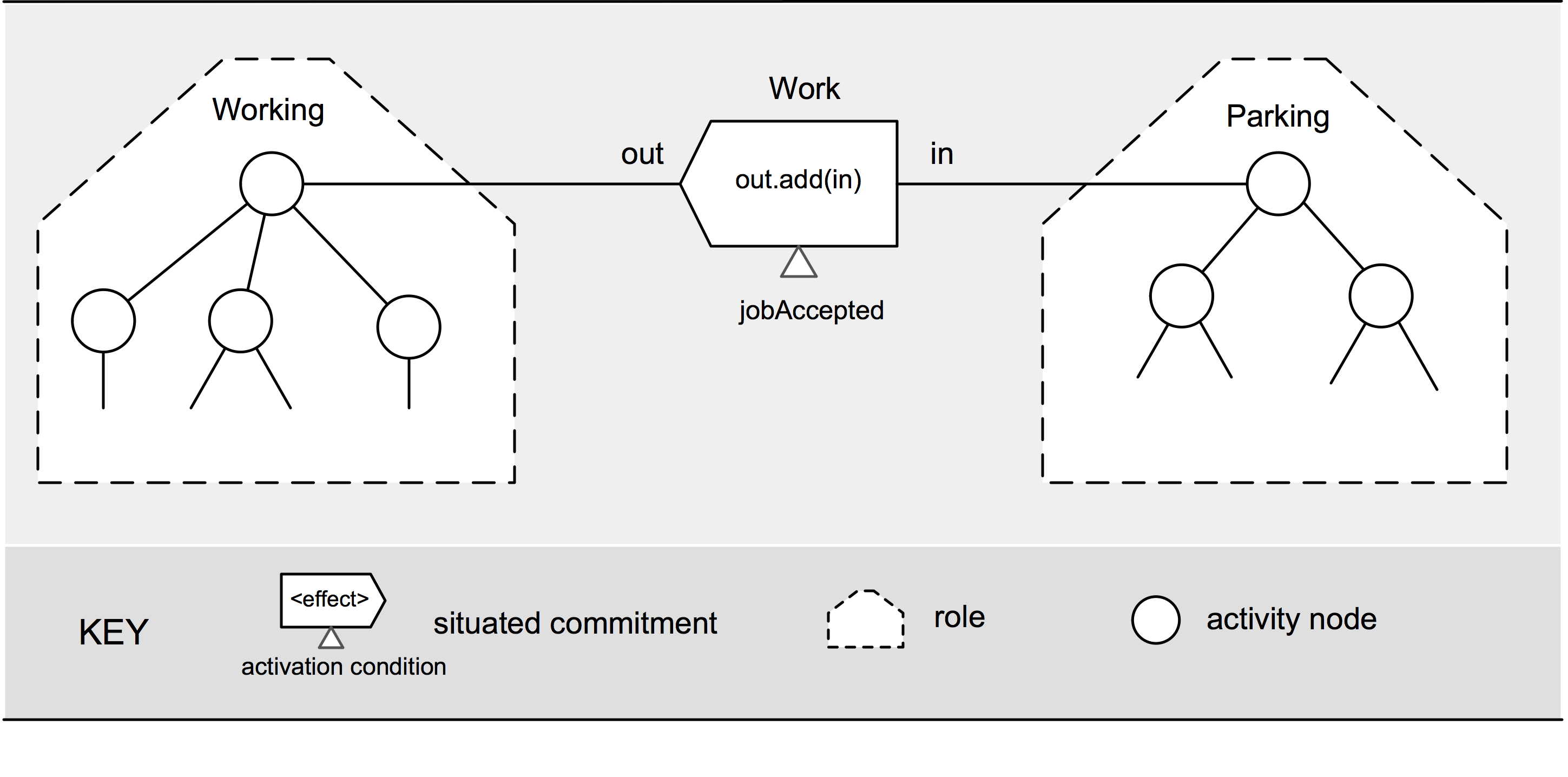}}
\caption{Design of the work commitment of AGV agents}
\label{fig:tree-commitment}
\end{figure}

Fig.~\ref{fig:tree-commitment} shows the design of the work commitment for AGV agents. 

A situated commitment in the AGV transportation system is represented by a connector between a set of source roles (in) and one target role (out). When activated, additional activity is injected to the target role, proportional to the activity levels of the source roles~\cite{Stee,2008OOPSLA}. The work commitment connects the parking role (single source role) with the working role (target role) and is activated when the AGV has accepted a job (these elements define the context and condition of the situated commitment).  When the work commitment is activated it injects an additional amount of activity in the working role (i.e., the activity of the top node of the parking role is added to the top node of the working role). This additional activity will stimulate the working role to start performing the committed task. 
\vspace{4pt}\\
\textbf{Decision Making} is defined as follows:

\footnotesize{
\begin{verbatim}
component DecisionMaking is abstraction()
{
   type Role is view[name : String, behaviors : Any, 
             actions : set(Action)].
   type SituatedCommitment is view[name : String, context : Any, 
             condition Boolean, effect : Any].
   type Action is view[agentId : String, actionName : String,
             actionParams : Any].
   ...

   port updateReadRoles is { 
         connection updateRoles is in(Knowledge). 
         connection readRoles is out(set(Roles)). }    
   port updateReadCommitments is { 
         connection updateCommitments is in(Knowledge). 
         connection readCommitments is out(set(SituatedCommitment)). }    
   port act is { connection selectAction is out(Action).}.
   ...

   externalActions is location(set(Action)). 
   selectedAction is Action. 

   behavior is {
     selectAction is function (roles : set(Role), 
         commitments : set(SituatedCommitment) : 
           Action { unobservable. }.
      externalAction is function (action : Action, 
         actions : set(Action)) :  
           Boolean { unobservable. }.

      unobservable.
      via updateRoles send roleKnowledge : Knowledge.
      via updateCommitments send commitmentKnowledge : Knowledge. 
      via readRoles receive roles : set(Roles). 
      via readCommitments receive commitments : set(SituatedCommitment).
      selectedAction := selectAction(roles, commitments). 
      if (externalAction(selectedAction, externalActions)) 
          then
              via selectAction send selectedAction.
              behavior().
          else
             unobservable.  
             behavior().
   }
}
\end{verbatim}
}
\normalsize

The decision making component has a repository with external actions that define effective actions that can be invoked in the virtual environment. Actions that are not included in this set are internal to the agent and may be used to update the agent's current knowledge. Decision making starts with updating the state of the roles and situated commitments based on the agent's current knowledge. Next, the roles and situated commitments are read and an action is selected. Subsequently, the component checks whether the selected action is an external action, and if so the action it is invoked.  We illustrate the realization of the action selection function for the two example applications. 

\begin{figure}[h!]
\centering
\resizebox{\textwidth}{!}
{\includegraphics{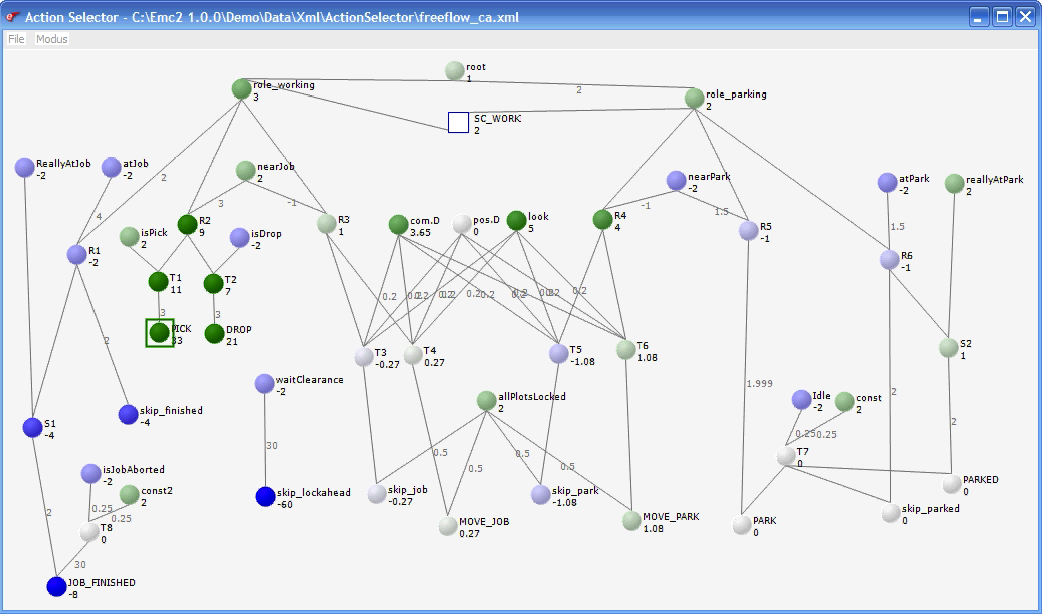}}
\caption{Runtime snapshot of the free-flow tree of an AGV agent. The part of the tree that deals with battery charging is omitted.}
\label{fig:tree-snapshot}
\end{figure}

We start with an excerpt of the decision making of an AGV agent that integrates the roles and situated commitments illustrated in Fig.~\ref{fig:tree-role} and Fig.~\ref{fig:tree-commitment}. To manage the complexity, we designed behavior-based action selection of the AGV agent in two subsequent steps.  In the first step, a free-flow tree is used to select an action at a high-level of abstraction. Examples of such actions are move, pick, and park. In the next step, the selected action is further refined taking into account collision avoidance and deadlock avoidance. For example, for a move action, the next segment is added if the corresponding path projection has been locked. We illustrate here the fist step, for additional information, we refer the interested reader to~\cite{Wey07,Wey09}. Fig.~\ref{fig:tree-snapshot} shows a snapshot of the main part of the free-flow tree of an AGV agent in action. 

The action selection function of the AGV agent is executed every second. To select an action, activity is injected via the root node of the tree. The root node feeds its activity to the working node and the parking node. These nodes feed the activity down through the hierarchy until the activity arrives at the action nodes, i.e. the leaf nodes of the tree. The level of activity that flows from one node to another may be multiplied by a weight factor that is associated with the edge. For example, the connection between the root node and the top node of the parking role has a weight factor of 2. The colors of the nodes in Fig.~\ref{fig:tree-snapshot} give a visual representation on the level of activities of the different parts of the behavior (the exact values are indicated next to each node). While the activity flows through the tree, some of the nodes receive additional activity from stimuli, e.g., the nodes just beneath the working role (R2 and R3) receive additional activity from the nearJob stimulus. When activated, the situated commitment work, that connects the parking role with the working role, injects an additional amount of activity in the working role (equal to the activity in the parking role). In the situation shown in Fig.~\ref{fig:tree-snapshot}, the situated commitment is active and as a result the activity level of the working role is 3, i.e., the sum of the activity 1 from the root node and 2 from the parking role via the situated commitment. When all action nodes have collected their activity the node with the highest activity level is selected for execution. In the example, the activity collected by the pick node is dominant (value 33) so the agent will select this action. The concrete node where to perform the action will be determined in the second step of action selection (see also the definition of the pick action in the discussion of the action service in section~\ref{sec:arch-elems-ve}). 

We conclude the discussion of the behavior-based action selection pattern with an excerpt of the design that illustrates the decision making of the vehicle agents in the anticipatory vehicle routing application. Figure~\ref{alg:vehicle:loop} shows the subsequent actions selected by the vehicle agent.
			
			\begin{figure}[!h]
				\begin{footnotesize}
				\algsetup{linenosize=\tiny}
				\renewcommand{\algorithmiccomment}[1]{\texttt{// #1}}
				\begin{algorithmic}[1]
					\LOOP
						\STATE \COMMENT{Update the feasible paths}
						\STATE $paths \leftarrow getAlternatives( currentLocation, destination )$\label{alg:vehicle:loop:beliefs}
						\STATE \COMMENT{Select the best route}
						\STATE $selectedPath \leftarrow choose( paths )$\label{alg:vehicle:loop:choose}
						\STATE \COMMENT{Decide whether to revise intention}
						\IF{$reviseIntention( currentIntention, selectedPath )$}\label{alg:vehicle:loop:revise}
						\STATE $currentIntention \leftarrow selectedPath$
						\ENDIF
						\STATE \COMMENT{Trigger propagation of the current intention}
						\STATE $propagateIntention( currentIntention )$\label{alg:vehicle:loop:propagate}
						\STATE \COMMENT{Instruct the driver}
						\STATE $instructDriver( currentIntention )$
					\ENDLOOP
				\end{algorithmic}
				\end{footnotesize}
				\caption{Decision making of a vehicle agent in the anticipatory vehicle routing system}
				\label{alg:vehicle:loop}
			\end{figure}

The behavior of the vehicle agent is contained in a single role, with the current intention (selected path) as a single type of situated commitment. On line~\ref{alg:vehicle:loop:beliefs}, the vehicle agent updates its knowledge with feasible paths explored by exploration ants.  On line~\ref{alg:vehicle:loop:choose}, the agent selects the shortest path based on the set of feasible paths. On line~\ref{alg:vehicle:loop:revise}, the vehicle agent decides on whether to deviate to this new path.  Whatever the agent decides, on line~\ref{alg:vehicle:loop:propagate}, the vehicle agent triggers sending out an intention ant to the infrastructure agents along the  path of its current intention. Finally, the driver is instructed about the current intention. The interested reader finds additional information in~~\cite{KULeuven-167666,1538945}.

\subsubsection{Design Rationale}\label{design-rationale-act-sel}
\mbox{\ }\vspace{-6pt}\newline

The primary principles that underlay the design of action selection are behavior-based decision making, role and situated commitment as first class concepts. 

Behavior based decision making enables an agent to act efficiently. A situated agent select its actions based on the locally perceived state of the environment and limited internal state. The reactive nature of behavior based decision making enables an agent to respond efficiently when the conditions in the system and the environment change. Nevertheless, locality and reactivity of decision making may affect that overall efficiency due to a lack of global and long term view on the system. Behavior based decision making is therefore suitable for interaction driven problem domains, rather than data driven domains.  This fits the target domain of the architecture style. 

The design of a behavior-based decision making mechanism may be a complex task for non-trivial agents. To manage this complexity, we introduced roles and situated commitments for situated agents. Roles and situated commitments provide design abstractions that allow to structure the agent behavior. Furthermore, roles and situated commitments provide the means for agents to set up collaborations in a particular context. A role represents a coherent part of functionality that can serve as a building block for social organization of an agent system. A situated commitment is defined based on the context in which agents are situated, which typically refers to their roles. Explicitly naming roles and situated commitments enables agents to set up collaborations, reflected in mutual commitments. Such collaboration can be set up by exchanging messages (see the Protocol-Based Communication pattern below). Once a collaboration is established, the mutual commitments will affect the selection of actions in favor of the agents' roles in the collaboration. The social behavior is typically driven by the context in which agents are situated (e.g., in field based transport assignment, agents are driven by the fields that are emitted in the virtual environment by the tasks). The latter results in high degree of openness (when a task or AGV enters the system, the agents take into account the new field, and when a task or AGV leaves the system the field simply disappears). 

\subsection{Protocol-Based Communication}

\subsubsection{Primary Presentation}
\mbox{\ }\vspace{-6pt}\newline

\begin{figure}[t!]
\centering
\resizebox{0.8\textwidth}{!}
{\includegraphics{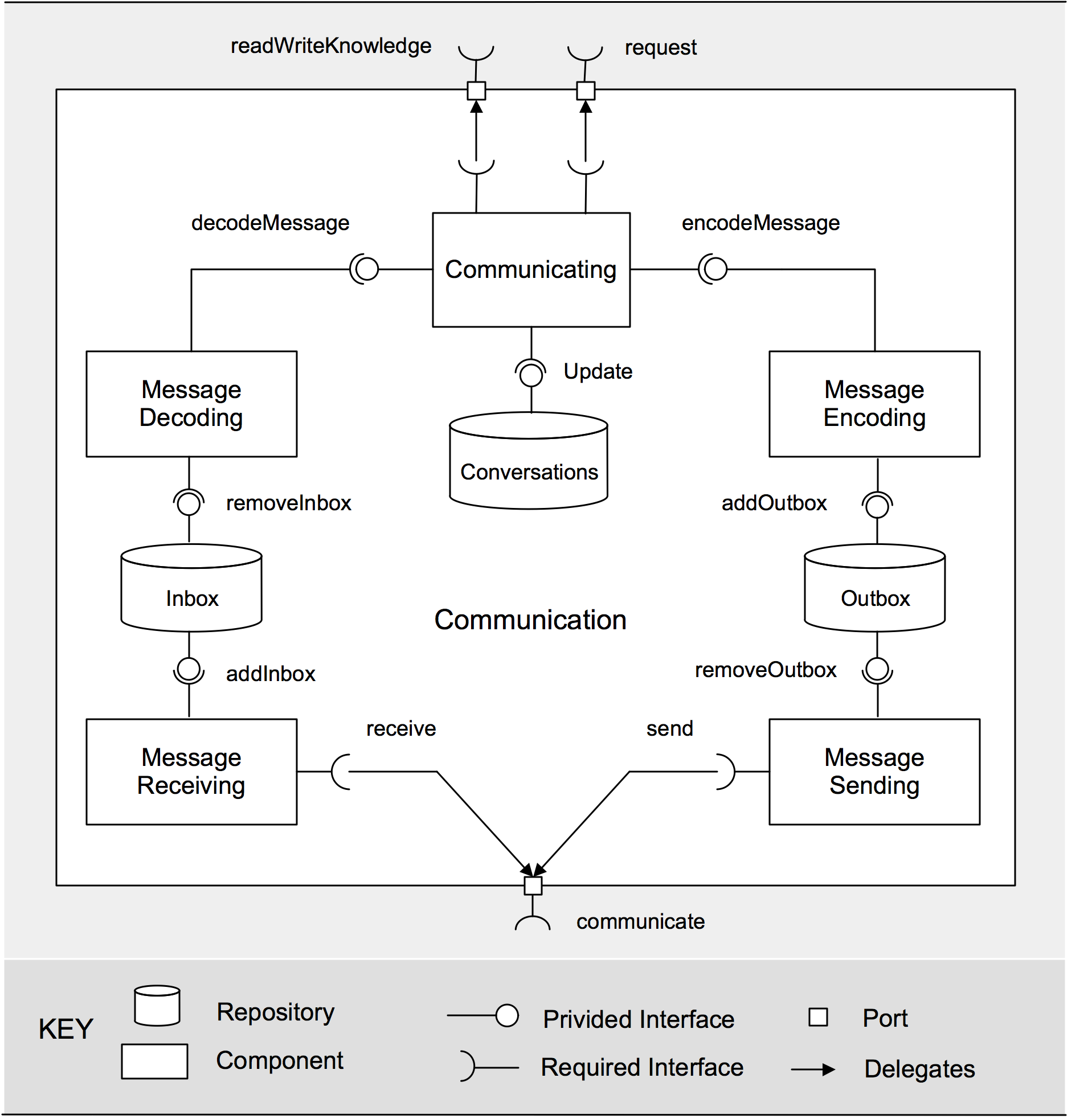}}
\caption{Protocol-based communication pattern}
\label{fig:protocol-based-communication}
\end{figure}

The primary presentation of the protocol-based communication pattern is shown in Fig.~\ref{fig:protocol-based-communication}.

Protocol-based communication enables situated agents to share information and set up collaborations. The communication component processes incoming messages, and produces outgoing messages according to well-defined communication protocols. A communication protocol specifies a set of possible sequences of messages.  The $\pi$-ADL specification of the communication component is defined as follows:

\footnotesize{
\begin{verbatim}
component Communication is abstraction()
{
   type Message is view[id : Integer, protocol_ : String, sender : String,
            protocol : String, performative : String, content : Any].
   ... 

   port communicate is {
          connection send is out(Message).
          connection receive is in(Message) }.
   ... 

   behavior is compose {
            messageReceiving is MessageReceiving().
      and   inbox is Inbox().
      and   messageDecoding is MessageDecoding().
      and   communicating is Communicating().
      and   conversatations is Conversations().
      and   messageEncoding is MessageEncoding().
      and   outbox is Outbox().
      and   messageSending is MessageSending().
   } where {
            receive relays messageReceiving::receive
      and   messageReceiving::addIn unifies inbox::addIn
      and   messageDecoding::removeIn unifies inbox::removeIn
      and   messageDecoding::decode unifies communicating::decode
      and   communicating::knowledgeTemplate relays knowledgeTemplate
      and   readKnowledge relays communicating::readKnowledge
      and   communicating::writeKnowledge relays writeKnowledge
      and   communicating::requestPerception relays requestPerception
      and   communicating::readUpdateConversation unifies
               conversations::readUpdateConversation
      and   communicating::encode unifies messageEncoding::encode
      and   messageEncoding::addOut unifies outbox::addOut
      and   messageSending::removeOut unifies outbox::removeOut
      and   messageSending::send relays send
      }                			
}
\end{verbatim}
}
\normalsize

Message receiving accepts messages from the communication service and adds them to the inbox message buffer. Message decoding selects messages from the inbox and decodes the messages. The communicating component interprets the decoded messages and updates the agent knowledge and conversations. Communicating also initiates and continues conversations when the necessary conditions hold (based on the agent's current knowledge and the ongoing conversations). Ongoing conversations are stored in the  conversations repository. A conversation is a communicative interaction following a well-defined communication protocol that defines of a series of protocol steps. Message encoding encodes newly composed messages and add them to the outbox message buffer. Finally, message sending selects pending messages from the outbox buffer and passes it to the communication service of the virtual environment.

\subsubsection{Architectural Elements}
\mbox{\ }\vspace{-6pt}\newline

We limit the discussion to the central components: message decoding, conversations, and communicating. Message encoding is basically the inverse of message decoding. The other components, message receiving, inbox, outbox, and message sending, have standard functions. The online Appendix A provides a rigorous definition of these components. 
\vspace{4pt}\\
\textbf{Message Decoding} is defined as follows: 

\footnotesize{
\begin{verbatim}
component MessageDecoding is abstraction()
{
   ...
   type Message is view[id : Integer, sender : String, receiver : String,
           protocol : String, performative : String, content : Any].
   type MessageData is view[id : Integer, sender : String, receiver : String,
           protocol : String, performative : String, content : sequence[Knowledge]].

   port removeInbox is { connection removeIn is in(Message) }.
   port decodeMessage is { connection decode is out(MessageData) }.

   language : Any.
   ontology : Any.  


   behavior is {
      decode is function( msg : Message, ontology : Any, language : Any ) :  
          MessageData { unobservable. }.

      unobservable.
      via removeIn receive message : Message.
      via decodeMsg send decode(message, language, ontology).
      behavior().
   }
}
\end{verbatim}
}
\normalsize

Message data is data extracted from a message in a form that can be interpret by the communicating component. Message decoding selects a received message from the inbox buffer and decodes the message according to the given content language and ontology. The content language defines the format of the messages, i.e.~the subsequent fields the message is composed of, and the ontology defines a shared vocabulary of words that agents use to represent domain concepts and relationships between the concepts. 

As an example, in the AGV transportation system, we defined a content language for the messages of the DynCNET protocol that allows tasks agents and AGV agents to agree on task assignment. 
DynCNET offers an alternative to field-based task assignment in the AGV transportation system, which we illustrated in section~\ref{sec:arch-elems-ve}. We elaborate on the protocol below. 
The content language defines for each performative a set of strings that represent fields of the corresponding messages. The ontology defines the domain of  values for each field. For example, the domain of the task priority field defines 5 priority levels represented as integers. During message decoding, the content of the message is parsed based on the performative of the message using the content language and the ontology. 
\vspace{4pt}\\
\textbf{Conversations} is defined as follows: 

\footnotesize{
\begin{verbatim}
component Conversations is abstraction()
{
   ...
   type MessageData is view[id : Integer, sender : String, receiver : String,
           protocol : String, performative : String, content : sequence[Knowledge]].
   type Conversation is sequence[MessageData].

   port readUpdateConversation is {
          connection selectConversation is in(Integer).
          connection readConversation is out(Converstation).
          connection updateConversation is in(MessageData).
          connection addConversation is in(MessageData).
          connection terminateConversation is in(Integer) }.

   type conversations is location(set(Conversation)).

   behavior is {
      add is function(data : MessageData, 
         convs : set(Conversation) ) :
          set(Conversation) { unobservable. }.
      write is function( id : Integer, 
         convs : set(Conversation) ) :
          set(Conversation) { unobservable. }.
      read is function( conv : Conversation,
        conv : Conversation ) :
           set(Conversation) { unobservable. }.
      terminate is function(id : Integer, 
        convs : set(Conversation) ) :
           set(Conversation) { unobservable. }.

      choose {
         //add conversation
        via addConversation receive data : MessageData.
        conversations := add(data, conversations). 
        behavior().
      or
         //read conversation
        via selectConversation receive id : Integer.
        via readConversation send readConversation(id, conversations).
        behavior().
      or
         //write conversation
         via updateConversation receive  data : MessageData.
         conversations := write(data, conversations). 
         behavior().
      or
         //delete conversation
         via terminateConversation receive id : Integer.
         conversations := terminate(id, conversations).
         behavior().
      or
     }
   }
}
\end{verbatim}
}
\normalsize

A conversation is a sequence of message data. The conversation component maintains a set of ongoing conversations. Conversations offers four functions that allows the communicating component to manage the conversations repository. Add includes a new conversation to the set of conversations, read returns a conversation based on a given id, write updates an ongoing conversation, and delete removes the conversation with a given id from the set of conversations. 
\vspace{4pt}\\
\textbf{Communicating}, the central component of the communication pattern, is defined as follows: 

\footnotesize{
\begin{verbatim}
component Communicating is abstraction ()
{
   type MessageData is view[id : Integer, sender : String, receiver : String,
           protocol : String, performative : String, content : sequence[Knowledge]].
   type Conversation is sequence[MessageData].
   type Protocol is view[name : String, steps : Any].
   ... 

   port readUpdateConversation is { ... }.
   port decodeMessage is { connection decode is in(MessageData) }.
   port encodeMessage is { connection encode is out(MessageData) }.
   ...

   protocols : set(Protocol).
   conversationIds : set(Integer).

   behavior is {
    extractKnowledge is function (data : MessageData) :  
          Knowledge { unobservable. }.
    addCommunication is function (data : MessageData) : 
          Knowledge { unobservable. }.
    updateCommunication is function (data : MessageData) : 
          Knowledge { unobservable. }.
    closeCommunication is function (c : Conversation) : 
          Knowledge { unobservable. }.
    ongoingConversation is function (id : Integer, ids : set(Integer)) :
          Boolen { unobservable. }.
    continueConversation is function (k : Knowledge, ids : set(Integer)) : 
          Boolen { unobservable. }.
    terminateConversation is function (ps : set(Protocol), 
       c: Conversation, k : Knowledge) : Boolean { unobservable. }.
    initiateConversation is function (ps : set(Protocol), k : Knowledge) :
          MessageData { unobservable. }.
    encodeMessage is function (ps : set(Protocol), c : Conversation, 
        k : Knowledge) : MessageData { unobservable. }.
    getId is function (k : Knowledge, ids : set(Integer)) : 
          Integer { unobservable. }.
    addId is function (id : Integer, ids : set(Integer)) :
          set(Integer) { unobservable. }.

     msg_knowledge : Knowledge. 
     enc_msg_data : Messagedata. 
     conv_id : Integer.

     choose
     {
         //handle incoming message
         via decode receive dec_msg_data : MessageData.
         msg_knowledge := extractKnowledge(dec_msg_data). 
         if(ongoingConversation(dec_msg_data::id, conversationIds)) 
            then {
              via updateConversation send dec_msg_data.  
              via writeKnowledge send updateCommuncation(dec_msg_data). 
            }
            else {  
              via addConversation send dec_msg_data.  
              via writeKnowledge send addCommuncation(dec_msg_data). 
            }
         behavior().
     }
     or
     { 
         //handle outgoing message
         unobservable. 
         via readKnowledge receive knowledge : Knowledge.        
         if(continueConversation(knowledge, conversationIds)) 
            then {
              via selectConversation send getId(knowledge, conversationIds).
              via readConversation receive conversation : Conversation.
              enc_msg_data := 
                   encodeMessage(protocols, conversation, knowledge). 
              via encode send enc_msg_data.
              via updateConversation send enc_msg_data.  
              via writeKnowledge send updateCommunication(enc_msg_data). 
           }
           else {  
              enc_msg_data := initiateConversation(protocols, knowledge).  
              via addConversation send enc_msg_data.  
              conversationIds := addId(id, conversationIds).
              via writeKnowledge send addCommunication(enc_msg_data). 
           }
         behavior().
     }
     or
     { 
         unobservable. 
         via readKnowledge receive knowledge : Knowledge.       
         conv_id := getId(knowledge, conversationIds).
         via selectConversation send conv_id.
         via readConversation receive conversation : Conversation. 
         if(terminiateConversation(protocols, conversation, knowledge)) 
            then {               
              //terminate conversation
              via terminateConversation send conv_id.
              conversationIds := deteleId(conv_id, conversationIds).
              via writeKnowledge send closeCommunication(conversation). 
            }
            else {  
              //silence
              unobservable.
            }
         behavior().
     }
   }
}
\end{verbatim}
}
\normalsize

\label{sec:arch-elems-ve}

Central to message-based communication are protocols. A protocol is characterized by a name and a set of protocol steps. To manage the communicative interactions, the communicating component maintains a set of conversation identifiers. The behavior of the communicating component consists of three parts: handling incoming messages, handling outgoing messages, and silent behavior. An incoming message can refer to an ongoing conversation or initiate a new conversation. An outgoing message can be the continuation of  an ongoing conversation or the initiation of a new conversation, depending on the conditions as reflected in the agent's current knowledge. Silent behavior can either terminate a conversation when the necessary conditions hold, or have no effects.

We illustrate protocol-based communication with DynCNET, a protocol for transport assignment that we developed for the AGV transportation system. 
DynCNET is an extension of the well-know contract net protocol Contract NET~\cite{ContractNet}, with ``Dyn'' referring to support for dynamic task assignment. The default protocol consists of four steps: (1) the initiator sends a call for proposals; (2) the participants respond with proposals; (3) the initiator notifies the provisional winner; and finally, (4) the selected participant informs the initiator that the task is started. These four steps are basically the same as in the standard Contract NET protocol.The flexibility of DynCNET is based on the provisional agreement between initiator and participant, and the possible revision of the assignment of the task between the third and fourth step of the protocol.

\begin{figure}[h!]
\begin{center}
\resizebox{.9\textwidth}{!}{\includegraphics{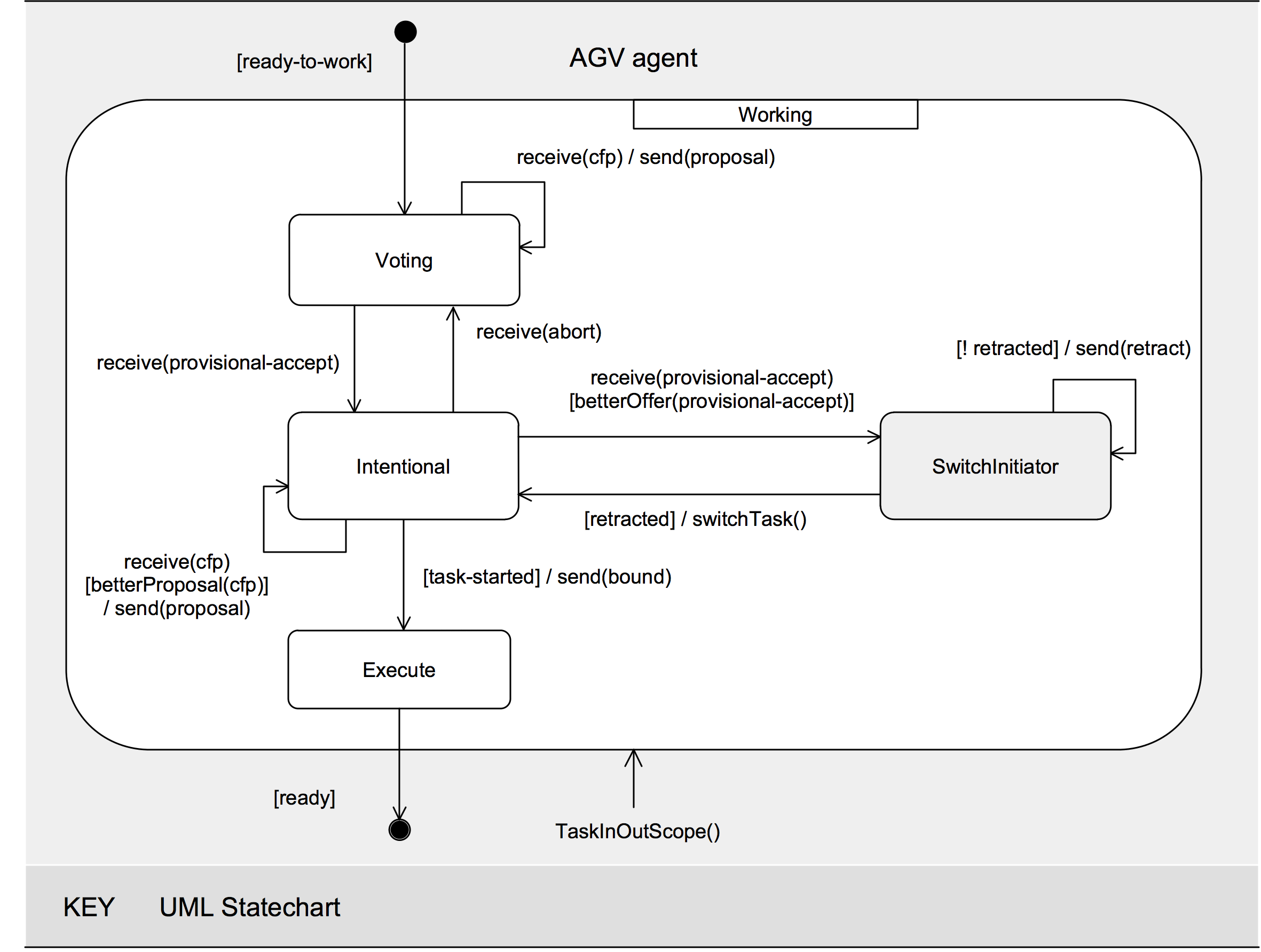}}
   \caption{DynCNET protocol for an AGV agent. In the shaded state, the agent can switch the provisional agreement. The format of a state transition is \emph{event [guard] / actions}.}
   \label{fig:protocolp}
\end{center}
\end{figure}

\begin{figure}[h!]
\begin{center}
\resizebox{.9\textwidth}{!}{\includegraphics{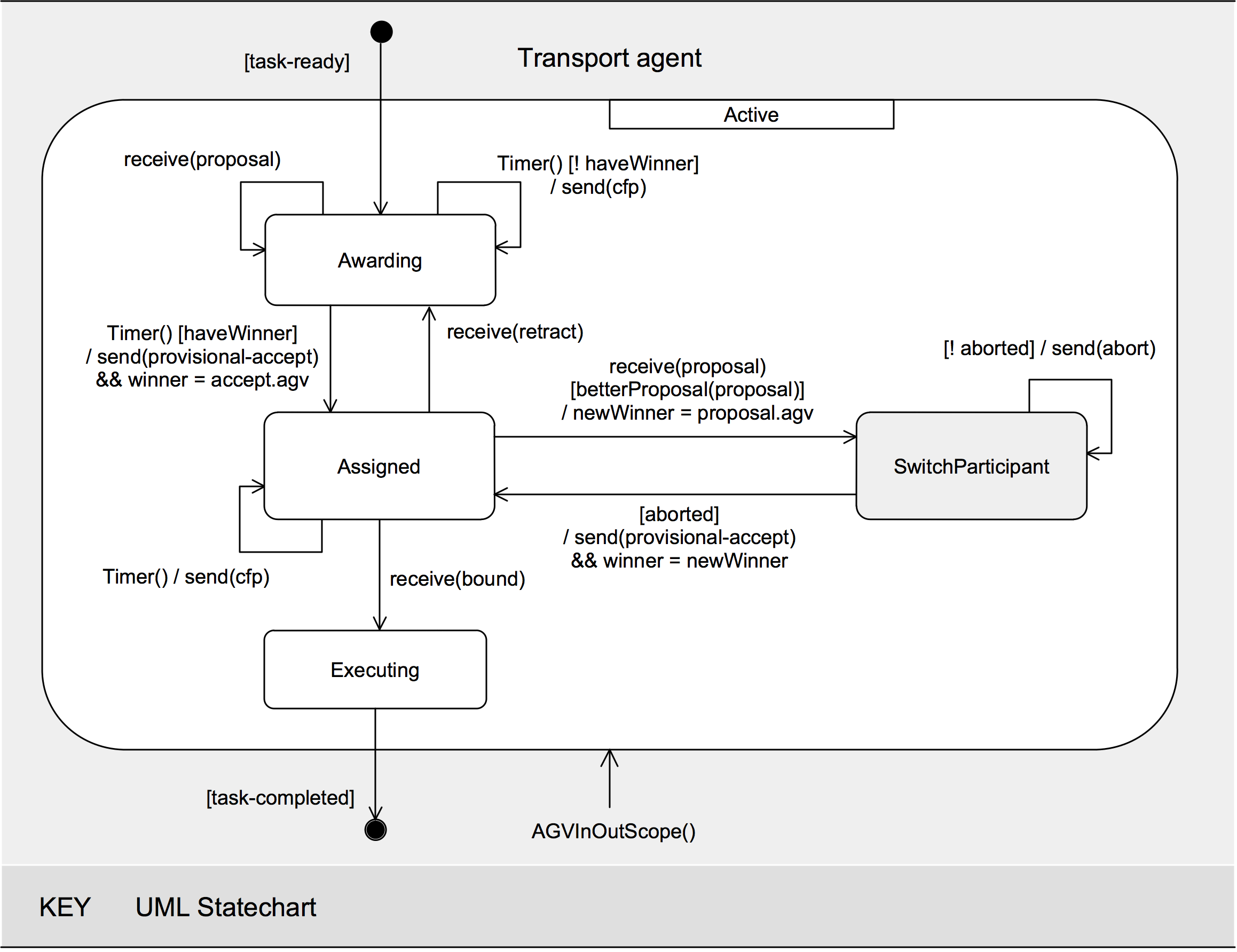}}
   \caption{DynCNET protocol for a transport agent. In the shaded state, the agent can switch the provisional agreement. The format of a state transition is \emph{event [guard] / actions}.}
   \label{fig:protocoli}
\end{center}
\end{figure}

These state diagrams in Fig.~\ref{fig:protocolp} and Fig.~\ref{fig:protocoli} show the behavior of the AGV agent and the transport agent respectively. When a task enters the system and it is ready to be executed (\texttt{task-ready}), the corresponding transport agent enters the \texttt{Active} state in which it remains until the task is completed (\texttt{task}--\texttt{completed}) (see Figure~\ref{fig:protocoli}). As soon as an AGV agent is \texttt{ready}--\texttt{to}--\texttt{work} it enters the \texttt{Working} state in which it remains until the task is executed (\texttt{ready}) (see Figure~\ref{fig:protocolp}). 

Let us first explain adaptation from the viewpoint of an AGV agent. When an AGV agent is ready to execute a task, it enters the \texttt{Voting} state. As long as the agent has not received a provisional accept, it answers \texttt{cfp}'s with \texttt{proposals}. When the AGV agent receives a \texttt{provisional}--\texttt{accept} message, it enters the \texttt{Intentional} state. As soon as the AGV starts the task (\texttt{task}--\texttt{started}), it sends a \texttt{bound} message to the transport agent to inform the latter that the execution of the task is started. The AGV agent is then committed to execute the task.\footnote{The transport agent's state changes from \texttt{Assigned} to \texttt{Executing} when it receives the bound message from the AGV agent (see Figure~\ref{fig:protocoli}).} However, if a new opportunity occurs before the task is started, i.e.~the AGV agent receives a \texttt{provisional}--\texttt{accept} which is a better offer, the agent changes to the \texttt{Switch Initiator} state. The agent then retracts from the earlier provisional task assignment (\texttt{send(retract)}), and switches to the more suitable task (\texttt{SwitchTask()}) entering again the \texttt{Intentional} state. AGV agents use the function \texttt{TaskInOutScope()} to determine whether new tasks enter and leave their area of interest (see Figure~\ref{fig:protocolp}). This function is realized by perception component. 

Let us know look at adaptation from the transport agent viewpoint. As long as the transport agent has not selected an AGV to execute the task (\texttt{!}\ \texttt{haveWinner}), it sends periodically (\texttt{Timer()}) \texttt{cfp}'s to the AGV agents in scope. Based on the received \texttt{proposals} from the AGV agents, it selects a winner, sends a \texttt{provisional}--\texttt{accept} message, and enters the \texttt{Assigned} state. As soon as the transport agent receives a \texttt{bound} message from the selected AGV agent, it enters the state \texttt{Executing} in which the task is effectively started. However, if a new opportunity occurs before the task is started, i.e.~the transport agent receives a \texttt{proposal} from an AGV agent  which is better than the current provisionally accepted proposal, the transport agent changes to the \texttt{Switch Participant} state. In this state the transport agent sends an \texttt{abort} message to the provisionally assigned AGV agent, and subsequently sends a \texttt{provisional}--\texttt{accept} message to switch to the more suitable AGV agent (\texttt{newWinner}). The function \texttt{AGVInOutScope()} notifies the transport agent when AGVs enter and leave its area of interest. This function is realized by the agent's perception component. 

\subsubsection{Design Rationale}\label{design-rationale-communication}
\mbox{\ }\vspace{-6pt}\newline

The primary principles that underlay the design of protocol-based communication are coordination through message exchange and structured interactions through protocols.  




Coordination through message exchange complements indirect coordination via marks in the virtual environment such as field-based coordination. Direct communication allows situated agents to exchange information and set up explicit collaborations. By exchanging messages, agents can agree to take up roles in a collaboration, reflected in mutual commitments. Such commitments steer the behavior of the agents towards the realization of shared objectives. Collaborations based on mutual commitments require a careful design as it creates explicit dependencies between agents.  

Communication defined in terms of protocols structures the communication, which fits the reactive nature of situated agents. In each step of a communicative interaction, conditions determine the agent's behavior in the conversation. Conditions not only depend on the status of the ongoing conversations and the content of received messages, but also on the actual conditions in the environment reflected in the agent's current knowledge. This contributes to the flexibility of the agent's behavior.



\section{Methodological Guidelines}\label{sec:guidelines}

We explained the key characteristics and requirements of the systems targeted by the architectural style in section~\ref{TargetDomain}. From our experience, we derive a set of guidelines to apply the style for systems within the target scope. Our objective is to offer the reader some guidance with respect to important aspects of the design of a situated multi-agent systems when using the architectural style. 
\vspace{5pt} \\
\textbf{What agents to select?} A key design decision of a situated multi-agent system is to select the type and number of agents in a system. There is no standard rule to select agents. However, as a rule of thumb, active entities that perform tasks in an autonomous way are primary candidates to become agents. In both the systems we used in this report, we have  selected two types of agents: one type that performs the actual tasks for the users (AGV agents that perform transportation tasks and vehicle agents that guide drivers), and another type that supports the first type if agent (task agents and infrastructure agents). In purely technical systems, agents are typically selected for processes that perform tasks, and resources that encapsulate some activities. In socio-technical systems, agents often represent social actors in the system (in addition to processes and resources). For example, in~\cite{Hae10}, we selected different types of situated agents in a supply chain that represent the different actors in a supply chain. 
\vspace{5pt} \\
\textbf{Responsibilities of agents and virtual environment.} Another key decision in the design of a situated multi-agent system is assigning responsibilities to agents and the virtual environment. In particular, for decisions regarding coordination, the designer often has a choice to either assign the responsibility to agents are assign it the virtual environment. The key underlying principle here is complexity management through separation of concerns. An excellent example that illustrates this aspect was the choice between the protocol-based and field-based approach to assign tasks in the AGV transportation systems. Using a protocol puts the burden of coordination on the agents. The task of the communication service is simply routing messages. Using the field-based approach puts the burden of spreading and maintaining the fields on the virtual environment. In this option, the agents can observe fields that guide them to a task. Finding a good balance that takes into account the different coordination tasks is not trivial.   
\vspace{5pt} \\
\textbf{Variant selection.} The patterns of the architectural style have several built-in variability mechanisms. Situated agents can or cannot make use of protocol-based communication. In the latter case, the communication component of situated agent and the communication service of the virtual environment can be omitted.  The perception service may or may not require direct access to the external environment. In the latter case, the synchronization component can be responsible to synchronize the state repository of the virtual environment with corresponding elements in the  environment. We applied this approach in the AGV transportation system. The state of the virtual environment can be coded in the same format as the knowledge used by the agents. For such a setting, the interpreting and descriptions components of agents can be omitted. Other variants are filters of the perception component, commitments of behavior-based action selection, and synchronization of the virtual environment. 
\vspace{5pt} \\
\textbf{Supporting middleware.} As any modern distributed software system, a situated multi-agent system requires common middleware service such as basic support for distribution, support for persistency, transactions, security, etc. Such services can be provided by common middleware that provides a basic design and deployment platform. Often, additional domain-specific services are required, such as communication services or dedicated services to interact with the external environment. The integration with this underlying middleware and supporting services is a crucial aspect in the design of a situated multi-agent system.

\section{Related Work}\label{sec:related}

Over the last decade, many patterns for multi-agent systems have been documented. To obtain a good overview of the state of the art on these patterns, we have performed a systematic literature review. The review covered the main publication venues of the field since 1998 resulting in 206 patterns documented in in 39 articles.\footnote{Publications from SpringerLink, ACM Digital Library, IEEE Explore, Science Direct, CiteseerX, Google Scholar were searched automatically. Publications of The Knowledge Engineering Review (KER), Transactions on Autonomous and Adaptive Systems (TAAS) and the Journal of Autonomous Agents and Multi-Agent Systems (JAAMAS) were searched manually.}  The list of articles with the patterns is attached in the online Appendix B. We limit the discussion here to a classification of the patterns that we derived from the study, and we explain how the patterns of the architectural style presented in this report fit in this classification. The study material and additional results are available via http://homepage.lnu.se/staff/daweaa/SLR-MASPatterns.htm

The objective of classifying design patterns for multi-agent systems is to provide an intellectual graspable overview of the huge space of existing patterns. The classification offers engineers a general picture of the pattern space of multi-agent systems, and helps those who are not familiar with the domain to get an easy jump-start to understand the pattern space. Several researchers have proposed classifications of design patterns for MAS, but these classifications are either bound to a specific catalog of patterns, or to an development methodology~\cite{Kendall1998,Aridor1998,Tahara1999,AIP,Oluyomi2006,Chella2010}. The classification presented in this report covers the full space of patterns for multi-agent systems as document at the time of writing. 

\begin{figure}[h!]
\begin{center}
\resizebox{\textwidth}{!}{\includegraphics{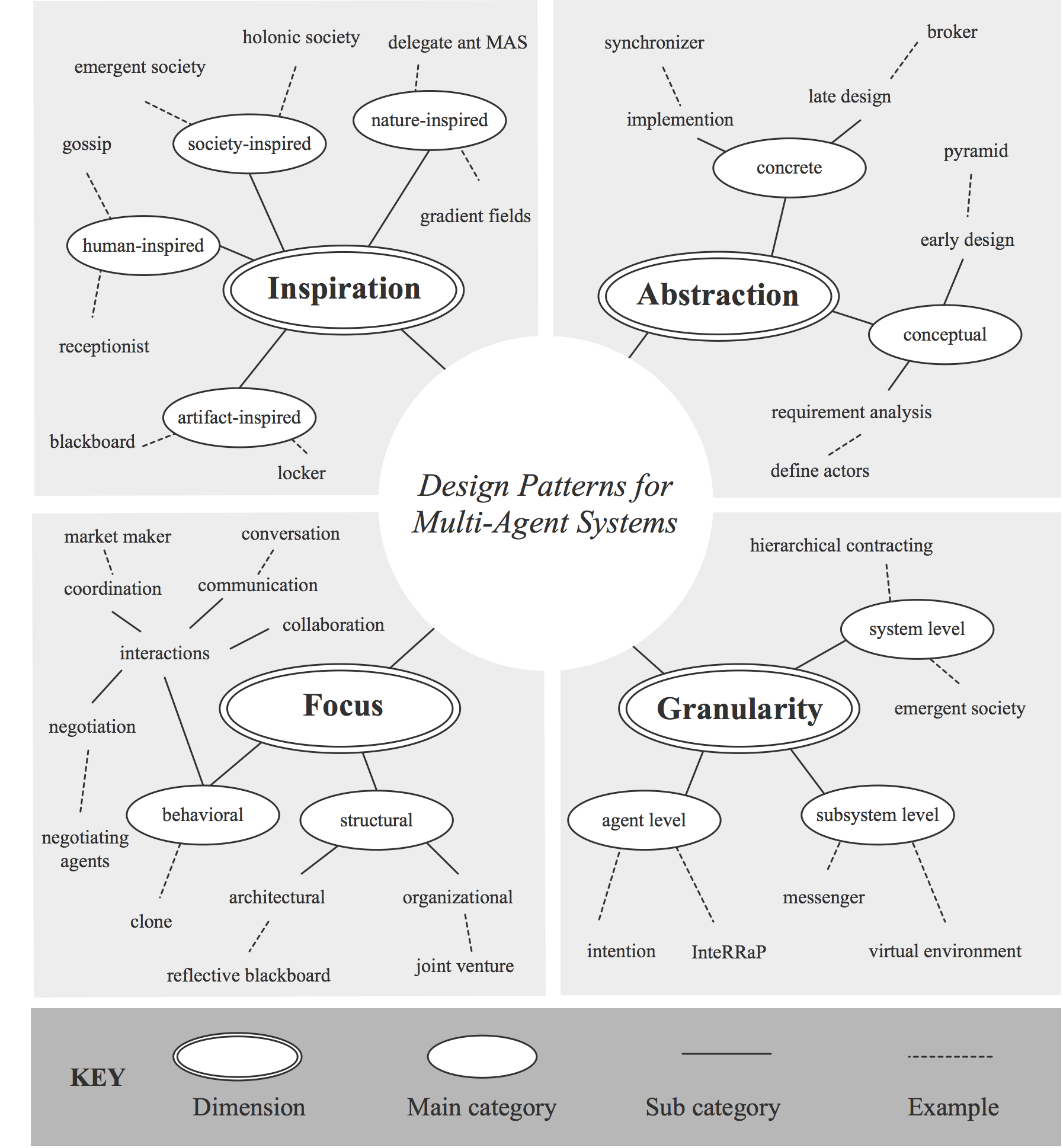}}
   \caption{Classification of patterns for multi-agent systems with example patterns. }
   \label{fig:patterns}
\end{center}
\end{figure}

The classification resulted from the analysis of several types of data that were collected during the review, including pattern categories (source of inspiration), pattern associations (explicitly documented relations between patterns), and short description of the patterns. From the analysis of the data, we derived four dimensions of patterns for multi-agent systems: inspiration, abstraction, focus, and granularity. Fig.~\ref{fig:patterns} shows a graphical overview of the dimensions, illustrated with example patterns.  

Metaphors and analogies help in understanding complex systems.  The \textit{inspiration} dimension has four categories that provide intuitive domains from which patterns are derived. Examples of nature-inspired patterns are gradient fields~\cite{DeWolf2007} that is inspired by the fields in nature, and delegate ant MAS~\cite{Holvoet:2009} that is inspired by behavior of social insects. Examples of society-inspired patterns are emergent society~\cite{Kendall1998} and holonic society~\cite{Chella2010} that get their inspiration from the way societies emerge and structure themselves. Examples of human-inspired patterns are receptionist~\cite{Meira2000} and gossip~\cite{Fernandez-Marquez2011}. Finally, locker~\cite{Aridor1998} and blackboard~\cite{Deugo1999} are example of patterns that get there inspiration from artifacts in our environment. 

The \textit{abstraction} dimension classifies patterns either as conceptual or concrete.  Both these main categories are further refined in subcategories that refer to stages in the software life cycle where the patterns can be used.  Actors~\cite{Mouratidis2006} and pyramid~\cite{Kolp2006} are examples of patterns that are useful in early phases in the life cycle, while broker~\cite{Hayden1999} and synchronizer~\cite{AIP} can be applied in detailed design and the implementation phase. 

The \textit{focus} dimension has two categories: structural and behavioral. Structural patterns are useful to deal with the decomposition of a system, while behavior patterns are useful to deal with interaction aspects. Examples of structural patterns are reflective blackboard~\cite{Silva2003} and  joint venture~\cite{Kolp2006}, the former focusing a particular coordination structure for an agent system, the latter focusing on the way a community of agents is organized. Examples of behavioral patterns are market maker~\cite{Deugo1999}, negotiating agents~\cite{Deugo1999}, and conversation~\cite{Kendall1998}. These patterns provide different approaches to support interactions among agents. 



Finally, the \textit{granularity} dimension refers to the scope of the patterns, i.e., the system or parts of the system.  Hierarchical contracting~\cite{Kolp2006} and emergent society~\cite{Kendall1998} are examples of patterns that apply to a multi-agent system as a whole.  Messenger~\cite{Aridor1998} and virtual environment~\cite{WeyWICSA2009} are patterns that apply to parts of a multi-agent system.  Intention~\cite{Kendall1998} and InteRRaP~\cite{Lind:2002} are patterns that support the design of individual agents.

 
We conclude with classifying the patterns presented in this report, see Table~\ref{tab:qs}. 

\begin{table}[!h]
\renewcommand{\arraystretch}{1.2}
\caption{Classification applied to the patterns of the architectural style} \label{tab:qs}
\begin{tabularx}{\columnwidth}{| l | l | X | X | l |}
\hline
\textbf{Pattern} & \textbf{Inspiration} & \textbf{Abstraction} & \textbf{Focus} & \textbf{Granularity} \\
\hline
Virtual environment & nature-inspired & conceptual (design) & structural \newline (architectural) & subsystem level \\
\hline
Situated agent & nature-inspired & conceptual (design) & structural \newline(architectural)  & agent level\\
\hline
Selective perception & artifact-inspired & conceptual (design) & behavioral & agent level\\
\hline
Behavior based action selection & society-inspired & conceptual (design) & behavioral & agent level\\
\hline
Protocol based communication & society-inspired & conceptual (design) & behavioral\newline (interactions) & agent level\\
\hline
\end{tabularx}
\end{table}

The patterns of the architectural style take inspiration from nature (environment, sensing, synchronization,e tc.), artifacts (focus, filter), and society (protocols, roles, situated commitments, etc.).  The five patterns are useful during architectural design. The primary patterns (virtual environment and situated agent) are structural, while the refined patterns are of behavioral nature. 
\section{Conclusions and Future Work}\label{sec:conclusions}

To further mature the domain of software architecture, \cite{Sha:Cle} argue for the creation of reference materials that give engineers access to the field's systematic knowledge. This report presents an architectural style for multi-agent systems that documents well-proven design expertise for a family of self-adaptive software systems. The structures and behaviors of the pattern elements are specified in $\pi$-ADL, a formal ADL supporting automated verification of architectural properties. These specifications provide rigorously defined templates to guide architects when applying the patterns in practice. The templates are under-specified providing the necessary flexibility for applying the patterns for a wide range of systems. We have illustrated how we have realized the various patterns of the style with excerpts from an experimental as well as an industrial case study. 

Whereas the presented architectural style consolidates design knowledge, expertise from building new systems brings new know-how that may be useful to extend or evolve the style. In recent work, we have built a number of applications in which the pattern realizations of the presented architecture style have been extended with additional features. In one extension, we have encapsulated the life-cycle management of agent organizations as a reusable service, which is modeled as an intermediate component between agent and virtual environment~\cite{MACODO-arch}. In another extension, we have added self-healing components to a situated agent system that deals with particular types of faults~\cite{Vromant:2011}. These self-healing components monitor agents and the virtual environment deployed on different nodes and make adaptations when faults occur. Both these extensions may eventually lead to two new patterns that extend the architectural style.  

To conclude, we hope that the systematic knowledge documented in this report will contribute to a better understanding of multi-agent system architectures and their value for designing complex self-adaptive systems.

\section*{Online Appendix} 

\small

\url{https://people.cs.kuleuven.be/danny.weyns/Appendix-1-2.pdf}

\normalsize

\section*{Acknowledgement}

The research presented in this report consolidates multiple collaborative research efforts spread over eight years. We acknowledge the valuable contributions of our colleagues in these efforts, including Kurt Schelfhout, Alexander Helleboogh, Nelis Boucke, Robrecht Haesevoets, Rutger Claes, Elke Steegmans, Bartosz Michalik, Tom Holvoet, Usman Iftikhar, Jan Wielemans, Tom Lefever, Rudy Vanhoutte, Wim Van Betsbrugge, Jan Vercammen, Zawar Qayyum, and Qin Xiong. 


%


\bibliographystyle{acmtrans}
\bibliography{patterns-paper-v2}

\end{document}